\documentclass[12pt, preprint, longabstract]{aastex}
\usepackage{ccaption,equation}
\usepackage[multidot]{grffile}
\usepackage[usenames]{color}
\usepackage[normalem]{ulem}
\newcommand\ltsima{$\; \buildrel <\over\sim \;$}
\newcommand\simlt{\lower.5ex\hbox{\ltsima}}
\newcommand\gtsima{$\; \buildrel >\over\sim \;$}
\newcommand\simgt{\lower.5ex\hbox{\gtsima}}

\shorttitle{The bulge microlensing optical depth and the stellar number count}
\shortauthors{Sumi \& Penny}

\begin{document}

\title{Possible Solution of the long-standing discrepancy in 
the Microlensing Optical Depth Toward the Galactic Bulge
by correcting the stellar number count
}

\author{
T.~Sumi\altaffilmark{1}, 
M.~T.~Penny\altaffilmark{2,3}
}

\altaffiltext{1}{Department of Earth and Space Science, Graduate School of Science, Osaka University, Toyonaka, Osaka 560-0043, Japan,\\
e-mail: {\tt sumi@ess.sci.osaka-u.ac.jp}}
\altaffiltext{2}
{Department of Astronomy, Ohio State University, 140 W. 18th Ave., Columbus, OH 43210, USA}
\altaffiltext{3}{Sagan Fellow}

\begin{abstract}
We find that significant incompleteness in stellar number counts results in a 
significant overestimate of the microlensing optical depth $\tau$ and event rate
per star per year $\Gamma$ toward the Galactic bulge from the first two years of the MOA-II survey. 
We find that the completeness in Red Clump Giant (RCG) counts $f_{\rm RC}$ 
decreases proportional to the galactic latitude $b$, as
$f_{\rm RC}=(0.63\pm0.11)-(0.052\pm0.028)\times b$, 
ranging between 1 and 0.7 at $b=-6^\circ\sim-1.5^\circ$.
The previous measurements  using all sources by Difference Image Analysis (DIA) 
by MACHO and MOA-I suffer the same bias.
On the other hand, the measurements using a RCG sample by OGLE-II, MACHO and EROS  were free from this bias 
because they selected only the events associated with the resolved stars.
Thus, the incompleteness both in the number of events and stellar number count cancel out.
We estimate $\tau$ and $\Gamma$ by correcting this incompleteness.
In the central fields with  $|l|<5^\circ$, we find
$\Gamma=[18.74\pm0.91]\times10^{-6}\exp[(0.53\pm0.05)(3-|b|)]$ \,star$^{-1}$\,{yr}$^{-1}$
and  $\tau_{200}=[1.84\pm0.14]\times10^{-6}\exp[(0.44\pm0.07)(3-|b|)]$
for the 427 events with $t_{\rm E}\leq200\,$days using all sources brighter than $I_s\leq20$ mag.
Our revised all-source $\tau$ measurements 
are about 2-$\sigma$ smaller than the other all-source measurements
and are consistent with the RCG measurements within 1-$\sigma$. 
We conclude that the long-standing problem on discrepancy between the high $\tau$ with all-source samples by DIA
and low $\tau$ with RCG samples can probably be explained by the incompleteness of the 
stellar number count. A model fit to these measurements predicts 
$\Gamma=4.60\pm0.25\times10^{-5}$\,star$^{-1}$\,{yr}$^{-1}$  
at $|b|\sim-1^\circ.4$ and $-2^\circ.25<l<3^\circ.75$ for sources with $I<20$,
where the future space mission WFIRST will observe.
\end{abstract}


\section{Introduction}
The gravitational microlensing optical depth and the event rate toward the Galactic Bulge (GB)
are known to be useful observables for the study of the stellar mass function and the structure and
kinematics of the Galaxy,
as these quantities, in addition to the microlensing timescale distribution
are related to the masses and velocities of lens objects (\citealt{pac91}, \citealt{gri91}, \citealt{novati2008}).
Currently, the microlensing survey groups:
MOA-II\footnotemark\footnotetext{\tt http://www.massey.ac.nz/\~{}iabond/alert/alert.html}, 
OGLE-IV\footnotemark\footnotetext{\tt http://www.astrouw.edu.pl/\~{}ogle/ogle4/ews/ews.html},
WiSE\footnotemark\footnotetext{\tt http://wise-obs.tau.ac.il/\~{}wingspan/} \citep{wise_survey}
and
KMTNet \citep{kmtnet} are detecting a couple of thousand of microlensing events every year
toward the GB.

The magnification of a microlensing event is described by
the minimum impact parameter ($u_{\rm 0}$) in units of Einstein radius $R_{\rm E}(M, D_{\rm s},D_{\rm l})$, 
the time of maximum magnification ($t_{0}$), 
the Einstein radius crossing time (or timescale) ($t_{\rm E}= R_{\rm E}/v_{\rm t}$), where
$v_{\rm t}$ is the transverse velocity of the lens relative to the line of sight,
$M$ is the lens mass, $D_{\rm s}$ and $D_{\rm l}$ are the distance to 
the source and the lens, respectively (\citealt{pac86}).

The microlensing optical depth, $\tau$ is the probability that any given source star 
is magnified by more than 1.34 (corresponding to the source being 
inside the Einstein ring disk of the lens) at any given time. 
This is directly related to the mass density of compact objects along the 
line of sight (\citealt{pac96}).
Theoretically, it is simpler than the 
microlensing event rate, because it doesn't depend on
the lens mass and lens-source relative velocity distribution. 
$\tau$ can be determined observationally from the following expression,
\begin{equation}
  \label{eq:opt}
  \tau = \frac{\pi}{2N_{\rm s}T_{\rm o}} \sum_i \frac{t_{{\rm E},i}}{\varepsilon (t_{{\rm E},i})}.
\end{equation}
where $N_{\rm s}$ is the total number of source stars monitored for microlensing, 
$T_{\rm o}$ is the duration
of the survey in days, $t_{{\rm E},i}$ is the Einstein radius crossing time for 
the $i$-th event, and $\varepsilon (t_{{\rm E},i})$
is the detection efficiency at that time-scale.
Because long $t_{\rm E}$ events give a large contribution to $\tau$, we present
the observed optical depth with a subscript, which indicates
the maximum $t_{\rm E}$ value allowed by each analysis.

The previous Galactic bulge microlensing optical depth results have been somewhat
controversial (see details in \citealt{sumi2013}).
The first measurements of the optical depth, $\tau_{100}\sim3.3 \times 10^{-6}$ by 
OGLE (\citealt{uda94}) and $\tau_{150}\sim3.9^{+1.8}_{-1.2} \times 10^{-6}$ 
by MACHO (\citealt{alc97}), were well above the predictions of $\tau\sim 5\times 10^{-7}$ (\citealt{pac91}; \citealt{gri91}) 
and  $\tau\sim 8.5\times 10^{-7}$ \citep{kir94}.
The later studies based on Difference Image Analysis (DIA), which is less 
sensitive to the systematics of blending in crowded fields, also 
found relatively high values of
$\tau_{150}\sim2.5 \times 10^{-6}$ (at $b\sim -3.^{\circ}5)$ by MACHO (\citealt{alc00b}) 
and MOA (\citealt{sumi03}).

To explain high optical depths, the
presence of a bar oriented along our line of sight to the GB have been suggested
(\citealt{pac94}; \citealt{zha95};\citealt{han03}; \citealt{zha96}; \citealt{pea98}; \citealt{gyu99}). 
But their predictions range over $\tau= 0.8 - 2.0 \times 10^{-6}$ and had difficulty 
explaining the observed high optical depths.

\cite{alc97} raised the possibility of a systematic bias in the 
optical depth measurement due to the degeneracy between
$t_{\rm E}$ and $u_{\rm 0}$ in relatively low signal-to-noise
ratio (S/N) events when the source base-line flux is unknown
due to blending (c.f. \citealt{woz97}; \citealt{han99}; 
\citealt{bon01}; \citealt{gou02}). 

\cite{pop01} proposed that optical depth may be estimated without any bias 
due to blending by using only events with bright source stars, 
such as Red Clump Giants (RCG), in which the blending might be negligible, rather than using all sources 
including the faint sources as in previous studies.
Except for one high value measured by \cite{alc97},  the other measurements 
based on events with bright sources resulted in lower optical depths when measured by EROS (\citealt{afo03}),
MACHO (\citealt{pop05}),
and  EROS \citep{ham06}. 

However, \citet{pop05} and \citet{ham06} realized that lensing of a fainter star that is 
unresolved from the bright star are common. 
But they also noted that this would increase the apparent
number of bright star events, while it also make $t_{\rm E}$ shorter 
and these two effects would nearly cancel, so that it does not cause 
large bias in the optical depth.  
\cite{sumi2006} and \cite{smith2007} confirmed this cancelation by image level simulations.
They also measured the bulge optical depth from
OGLE-II for RCG sources with high S/N light curves, which allowed them to determine 
the source brightness and exclude events with faint sources. Thus, they do not
rely on the lucky cancelation of the biases.
Their result was consistent with the MACHO and EROS values.
These values are consistent with predictions based on the revised COBE bar model 
by \cite{han95}, which has a mass of $M_{\rm bulge}=1.62\times 10^{10} M_\odot$ 
and the viewing angle $\phi\sim 20^\circ$, and the latest COBE elongated bar 
model by \cite{bis02} with $\phi\sim 20^\circ$.

Although the optical depth difference between the RCG sources and the all
sources by DIA are not very significant due their large errors, the
DIA optical depth values are systematically larger than the RCG values. The reason
for this is not well understood. 

\cite{sumi2013} made optical depth measurements 
using samples of 83 RCG events and 474 all source events with well measured parameters from the DIA analysis of MOA-II data.
This is the largest sample ever used for an optical depth measurement.
Their optical depth measurement for all sources was in-between those of previous 
measurements, i.e., lower than all source samples and higher than RCG samples
and concluded that previous discrepancy between all source sample and RCG sample 
were just a statistical fluctuation.

\cite{sumi2013} pointed a possible problem on using the same luminosity 
function in all fields as the one in Baade's window for estimating the number of sources at small level.
But they did not consider the completeness of stellar number counts used to normalize 
that luminosity function. A systematic bias in the number counts of source stars will affect the measured optical depth. We investigate this point in this paper.

The event rate per star per year $\Gamma$ is also affected by the same bias.
This is important for the future space-based microlensing
surveys \citep{ben02}, like the exoplanet microlensing survey planned for
WFIRST \citep{green2012,Spergel2015} or Euclid \citep{penny13}

In this paper we estimate the completeness of the number count of the source stars
and revise the measurement of the microlensing event rate and
optical depth toward the GB based on the first two years of the MOA-II survey. 
We present the stellar number count in section \S\,\ref{sec:number_count}
and its completeness in section \S\,\ref{sec:completeness}.
We present the revised event rate and optical depth results in section \S\,\ref{sec:opt}. 
In section \S\,\ref{sec:model_opt} and  \S\,\ref{sec:model_gamma} , we model the distribution of the
optical depth and  event rate with galactic coordinates. The discussion and
conclusions are given in section \S\,\ref{sec:summary},

\section{Stellar number count.}
\label{sec:number_count}

\subsection{MOA Stellar number count}

We use the same dataset as \cite{sumi2013} which used the data taken in the 2006 and 2007 seasons by the MOA-II survey,
with the 1.8-m MOA-II telescope located at the Mt.\ John University
Observatory, New Zealand.
The telescope is equipped with the mosaic CCD camera, MOA-cam3  \citep{sako2008}, which 
has a 2.18 deg$^2$ field of view (FOV) with a pixel scale of 0.58 arcsec/pixel.
The median seeing for this dataset was $\sim2.0''$. 

The centers of the 22 GB fields of the MOA-II survey are listed in Table~\ref{tbl:fld}.
The images were taken using the custom MOA-Red wide-band filter, which
is equivalent to the sum of the standard Kron/Cousins $R$ and $I$-bands.
The average instrumental magnitudes of the MOA reference images were roughly
calibrated to the Kron/Cousins $I$-band using OGLE-II photometry map of the
Galactic bulge \citep{uda02} within $\sim$0.2 mag.  $V$-band images were taken
occasionally in order to make instrumental color-magnitude diagrams (CMD).

The images were reduced with MOA's implementation \citep{bon01} of 
the difference image analysis (DIA) method \citep{tom96,ala98,ala00}. 
In the DIA method, a high-quality,
good-seeing reference image is subtracted from each observed
image after matching the seeing and photometric scaling. 
This method provides precise relative photometry in very crowded stellar fields.
A stellar catalog was constructed from these reference images by applying DoPHOT \citep{sch93},
the point spread function (PSF)-fitting routine.

Each field is divided into 80 subfields and each subfield is individually
calibrated using the RCG feature in each subfield CMD more precisely.
About 12\% of the area, in which a clear RCG population could not
be identified in the CMD, was excluded from the analysis.
The number of subfields used in the final analysis is 1536 in total and also 
given in Table~\ref{tbl:fld} for each field.
The coordinates and other properties of the subfields are listed in  Table~\ref{tbl:opt_2D}.

For the microlensing rate and optical depth estimates,
we use two subsamples of events and star counts:

(1) The all-source sample uses stars brighter than $I_s \leq 20$ mag.
This sample contains 474 events.
\citep{sumi2013} did not require that the events  be associated with an apparently resolved 
reference image star, but did require that the source magnitude is determined from the light curve fit
and it is brighter than  $I_s = 20$ mag.
Analysis of these samples is less affected by blending, in the same way as previous all source DIA analyses \citep{sumi03,alc00b}.
However, the analysis requires that the number of sources be counted independently from the event selection.

The GB fields are so crowded that virtually all the main sequence stars
are not individually resolved. 
To count the number of stars with $I\le 20$, \citet{sumi2013} first estimated the center of RCG $I$-band magnitude, $I_{\rm RC}$, 
and the number of RCG, $N_{\rm RC}$, by fitting the magnitude distribution of the reference 
images in each subfield with Equation (4) of \cite{Nataf2013}. RCG stars are abundant and serve as a good standard candle (\citealt{kir97,sta00}) that trace out the density structure of the GB, and hence their numbers should be proportional to the number of all sources.

\citet{sumi2013} then constructed a combined luminosity function (LF) by using the star catalogs measured in Baade's Window 
using the MOA-II reference image for bright stars, and {\it Hubble Space Telescope} (HST) 
imaging \citep{hol98} for faint stars down to $I = 24$ mag.
This combined LF is calibrated to the extinction and GB distance, and normalized for
each subfield so that its $I_{\rm RC}$ and $N_{\rm RC}$ are same as the 
values in each subfield. Then the number of stars $N_{\rm s}$ are counted down to $I=20$ mag by 
integrating this scaled-combined LF as shown in Table~\ref{tbl:fld}.

The disadvantage of this method is that it assumes that the LF in all fields is the same as that of Baade's window \citep{hol98}.
The advantage of this method is that it was believed that faint sources can be counted without any problem with blending, 
because $N_{\rm RC}$ is less affected by  blending because RCGs are bright. 
However, below, we show that $N_{\rm RC}$ does suffer from incompleteness and that this method
is not tolerant against this bias for the measurements of $\tau$ and $\Gamma$ because 
the number of events is not affected by this incompleteness.

(2) The Red Clump Giant (RCG) sample selects only events with $I_{\rm s}<17.5$, as 
measured from the lightcurve. To estimate the number of sources, stars in the 
"extended RCG region" are counted in the CMD of the reference images as shown 
in Figure 1 of \citep{sumi2013} with $I_{\rm s}<17.5$ mag
and the source colors of $(V-I)_s\geq(V-I)_{\rm RC}-0.3$ mag,
where $(V-I)_{\rm RC}$ is the $V-I$ color of RCG centroid. There is no color cut on the event selection, but it is assumed that the blue disk sources in front 
of the bulge have negligible event rate.
This process is similar to the OGLE-II optical depth analysis \citep{sumi2006} that makes use of the OGLE-II 
extinction map \citep{sumEX04}, which is based on the RCG position in the CMD.
This contains not only RCGs but also bulge red giants, which is a similar definition 
 to previous works \citep{alc97,pop05,sumi2006,ham06}.
This sample contains 83 events.

Contrary to previous RCG analyses \citep{alc97,pop05,sumi2006,ham06},
\citet{sumi2013} did not require the event to be associated with an apparently resolved star. 
So this method is closer to the all source sample analysis above than the previous RCG analyses.
Thus their MOA-II RCG analysis is less affected by blending, but
affected by the same incompleteness bias as the all-source analysis. 
On the other hand, incompleteness of the source star count did not affect the previous RCG analyses 
because the incompleteness in event selection and source count cancel each other out 
(see more details in \S\,\ref{sec:summary}).

\subsection{OGLE Stellar number count}
\label{sec:OGLE}

The Optical Gravitational Lensing Experiment (OGLE; \citealt{uda03}) also
conducts a microlensing survey toward the Galactic bulge with the 1.3 m Warsaw telescope
at the Las Campanas Observatory in Chile.  The median seeing is about 1.3 arcsec. 
The third phase of OGLE, OGLE-III carried out survey observations with a 0.36 deg$^2$ FOV mosaic CCD camera. 
Most observations are taken in the standard Kron-Cousin $I$-band with 
occasional observations in the Johnson $V$-band.

\citet{Nataf2013} identified RCGs in the CMDs by using OGLE-III photometry maps towards 
the galactic bulge fields \citep{Szymanski2011} \footnote[3]{http://ogle.astrouw.edu.pl/} which 
cover $-10^\circ < l< 10^\circ$ and $2^\circ < |b| <7^\circ $.
Each of the 2104 OGLE-III subfields (eight detectors over 263 fields) used in 
the work was split into 1, 2, 3, 6, 8, 10, 15, or 21 rectangles
depending on the surface density of stars.  The average rectangle size is $6'\times 6'$.
In each rectangle,
they estimated the center of the RCG $I$-band magnitude, $I_{\rm RC}$, 
and the number of RCG, $N_{\rm RC}$ by fitting the luminosity function with Equation (4) of \cite{Nataf2013}.

Thanks to their better seeing and longer exposure than MOA-II, the completeness of the OGLE-III RCG number count 
is much higher than that of the MOA-II catalog and likely to be almost complete.

\section{Completeness of the Stellar Number Count}
\label{sec:completeness}

\subsection{Comparison to the OGLE RCG number count}

We investigate the completeness of the number count of RCGs in the MOA-II GB fields used 
in \cite{sumi2013}, $N_{\rm RC,MOA}$,
by comparing it to that of OGLE-III \citep{Nataf2013}, $N_{\rm RC, Nataf}$.

\begin{figure}
\begin{center}
\includegraphics[angle=-90,scale=0.7,keepaspectratio]{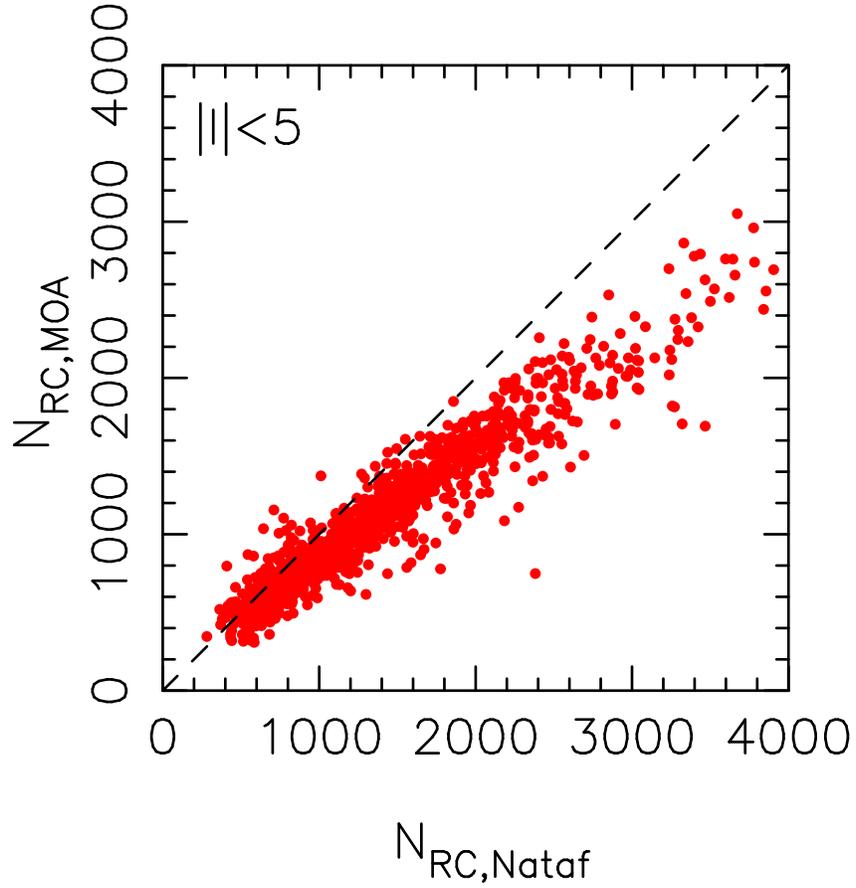}
\caption{
  \label{fig:NRC_Nataf_MOA}
The comparison of the number of red clump giants (RCG) per subfield (98 min.$^2$) in MOA, $N_{\rm RC,MOA}$  and
the that in OGLE, $N_{\rm RC,Nataf}$, which are  an average over points within 0.085 degrees of the MOA subfield center.
}
\end{center}
\end{figure}

Figure \ref{fig:NRC_Nataf_MOA}  shows comparison of 
the number of RCG per subfield (98 arcmin$^2$) in MOA ($N_{\rm RC,MOA}$) at $|l|<5^\circ$ and
 that in OGLE ($N_{\rm RC,Nataf}$) which are  an average over points within 0.085 degrees of the MOA subfield center.
One can see that they are consistent at low number density $N_{\rm RC}<1000$, 
but as the stellar density increases, $N_{\rm RC,MOA}$ becomes systematically lower by up to 30\%.
This trend is as expected because the completeness depends on the stellar number density, but the magnitude of the difference is 
significantly larger than anticipated by previous studies, and this could bias the measured $\tau$ and $\Gamma$.

\begin{figure}
\begin{center}
\includegraphics[angle=-90,scale=0.7,keepaspectratio]{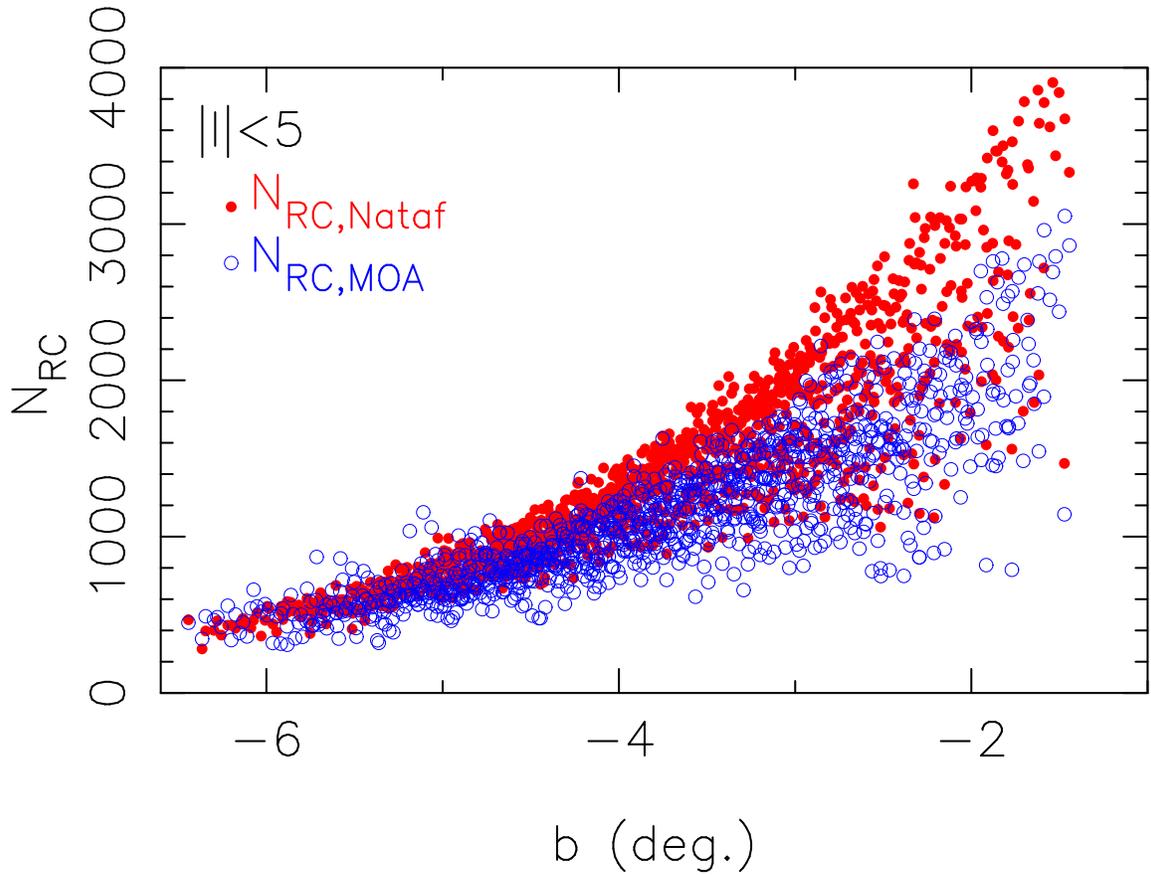}
\caption{
  \label{fig:NRC_b}
The number of RCG per subfield (98 arcmin$^2$) in MOA ($N_{\rm RC,MOA}$, blue open circle)  and
that in OGLE ($N_{\rm RC,Nataf}$, red filled circle) which are  an average over points within 0.085 degrees of the MOA subfield center,
as a function of the galactic latitude $b$.
}
\end{center}
\end{figure}

Figure \ref{fig:NRC_b} shows the counts as a function of galactic latitude $b$.
One can see that they are consistent at higher galactic latitude around $b\sim-6$, but $N_{\rm RC,MOA}$ is systematically fewer than $N_{\rm RC, Nataf}$ at 
lower $b$ as the number density is higher near the galactic center.
 
\begin{figure}
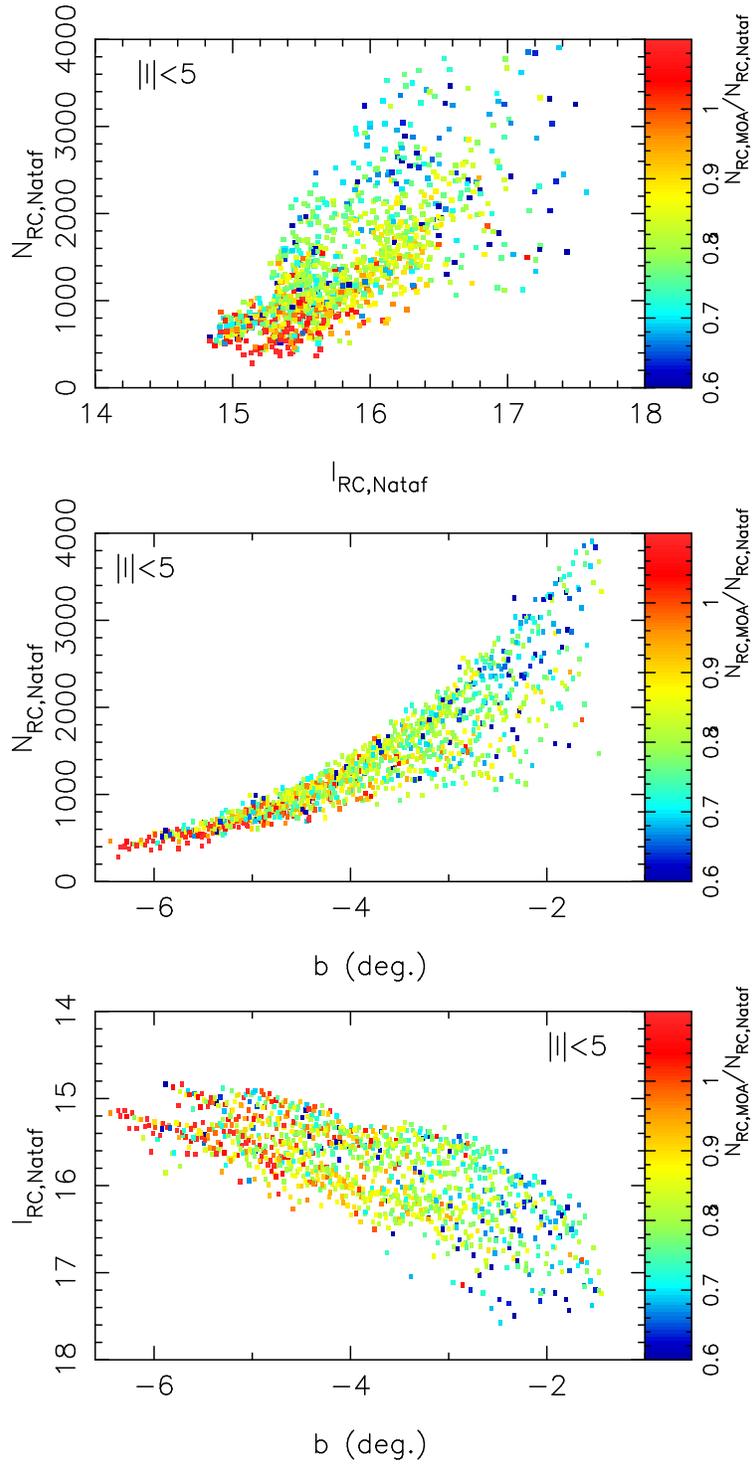

\begin{center}
\includegraphics[angle=-90,scale=0.39,keepaspectratio]{Figure3a.eps}
\includegraphics[angle=-90,scale=0.39,keepaspectratio]{Figure3b.eps}
\includegraphics[angle=-90,scale=0.39,keepaspectratio]{Figure3c.eps}
\caption{
  \label{fig:fRC_color_map}
The number of RCG measured by MOA over that measured by OGLE, $f_{\rm RC}=N_{\rm RC,MOA}/N_{\rm RC,Nataf}$
(color-coded) of subfields with $|l|\le5^\circ$  in  ($I_{\rm RC,Nataf}$, $N_{\rm RC,Nataf}$) (top panel),  ($b$, $N_{\rm RC,Nataf}$) (midle panel)
and ($b$, $I_{\rm RC,Nataf}$) (bottom panel) planes.
 }
\end{center}
\end{figure}

The completeness of RCG counts is expected to depend not only on the number density, but also the RCG magnitude,  
which depends on the interstellar extinction and the distance to the galactic bar structure.
The top panel of Figure \ref{fig:fRC_color_map} shows the ratio, $f_{\rm RC}=N_{\rm RC,MOA}/N_{\rm RC,Nataf}$
i.e., the completeness if we assume that $N_{\rm RC, Nataf}$ is complete, as a function of the $I$-band RCG magnitude measured by \cite{Nataf2013}, $I_{\rm Nataf}$ and of $N_{\rm RC,Nataf}$.
One can see that $f_{\rm RC}$ is basically higher for brighter $I_{\rm Nataf}$ and smaller $N_{\rm RC,Nataf}$ as expected,
but the trend is somewhat complicated. 
The middle and bottom panels of Figure \ref{fig:fRC_color_map} show $f_{\rm RC}$ in 
($b$, $N_{\rm RC,Nataf}$) and ($b$, $I_{\rm RC,Nataf}$) space, respectively. 
Both  $N_{\rm RC,Nataf}$ and $I_{\rm RC,Nataf}$ have a clear relation with $b$, which explains the
clear trend of $N_{\rm RC}$s with $b$ in Figure \ref{fig:NRC_b}.
In the top and middle panels of Figure \ref{fig:fRC_color_map}, one can see that the $f_{\rm RC}$
are systematically higher at smaller $N_{\rm RC,Nataf}$ at given $I_{\rm RC,Nataf}$ and $b$.
Some fraction of this trend can be attributed to the bias due to the statistical uncertainty
of $N_{\rm RC,Nataf}$ itself.
Because $f_{\rm RC}$ is inversely proportional to $N_{\rm RC,Nataf}$, 
$f_{\rm RC}$ correlates with $N_{\rm RC,Nataf}$.

\begin{figure}
\begin{center}
\includegraphics[angle=-90,scale=0.5,keepaspectratio]{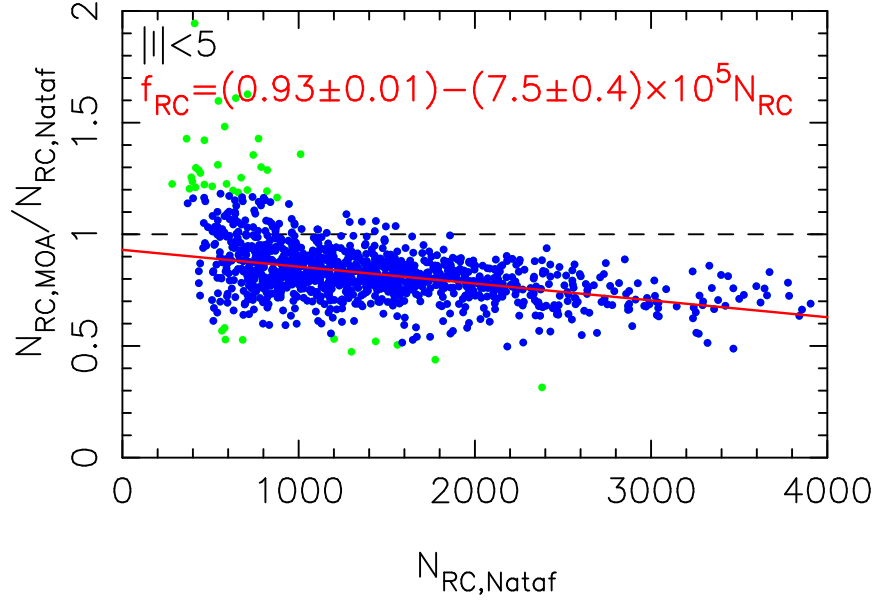}
\caption{
  \label{fig:fRC_NRC}
The number of RCG measured by MOA over that measured by OGLE, $f_{\rm RC}=N_{\rm RC,MOA}/N_{\rm RC,Nataf}$
for subfields with $|l|\le5^\circ$ as a function of $N_{\rm RC,Nataf}$.
The red line indicates the best fit to the blue dots where 
3$\sigma$ outliers (green dots) are recursively rejected. 
}
\end{center}
\end{figure}

\begin{figure}
\begin{center}
\includegraphics[angle=-90,scale=0.5,keepaspectratio]{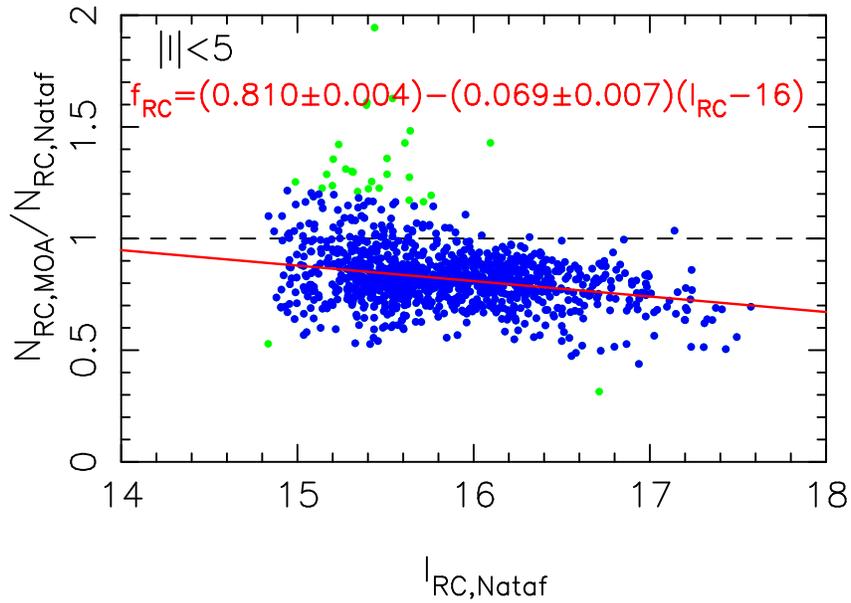}
\caption{
  \label{fig:fRC_IRC}
The number of RCG by MOA over that by OGLE, $f_{\rm RC}=N_{\rm RC,MOA}/N_{\rm RC,Nataf}$
of subfields with $|l|\le5^\circ$ as a function of $I_{\rm RC,Nataf}$.
The red lines indicate the best fit by using the blue dots where 
3$\sigma$ outliers (green dots) are recursively rejected. 
}
\end{center}
\end{figure}

\begin{figure}
\begin{center}
\includegraphics[angle=-90,scale=0.7,keepaspectratio]{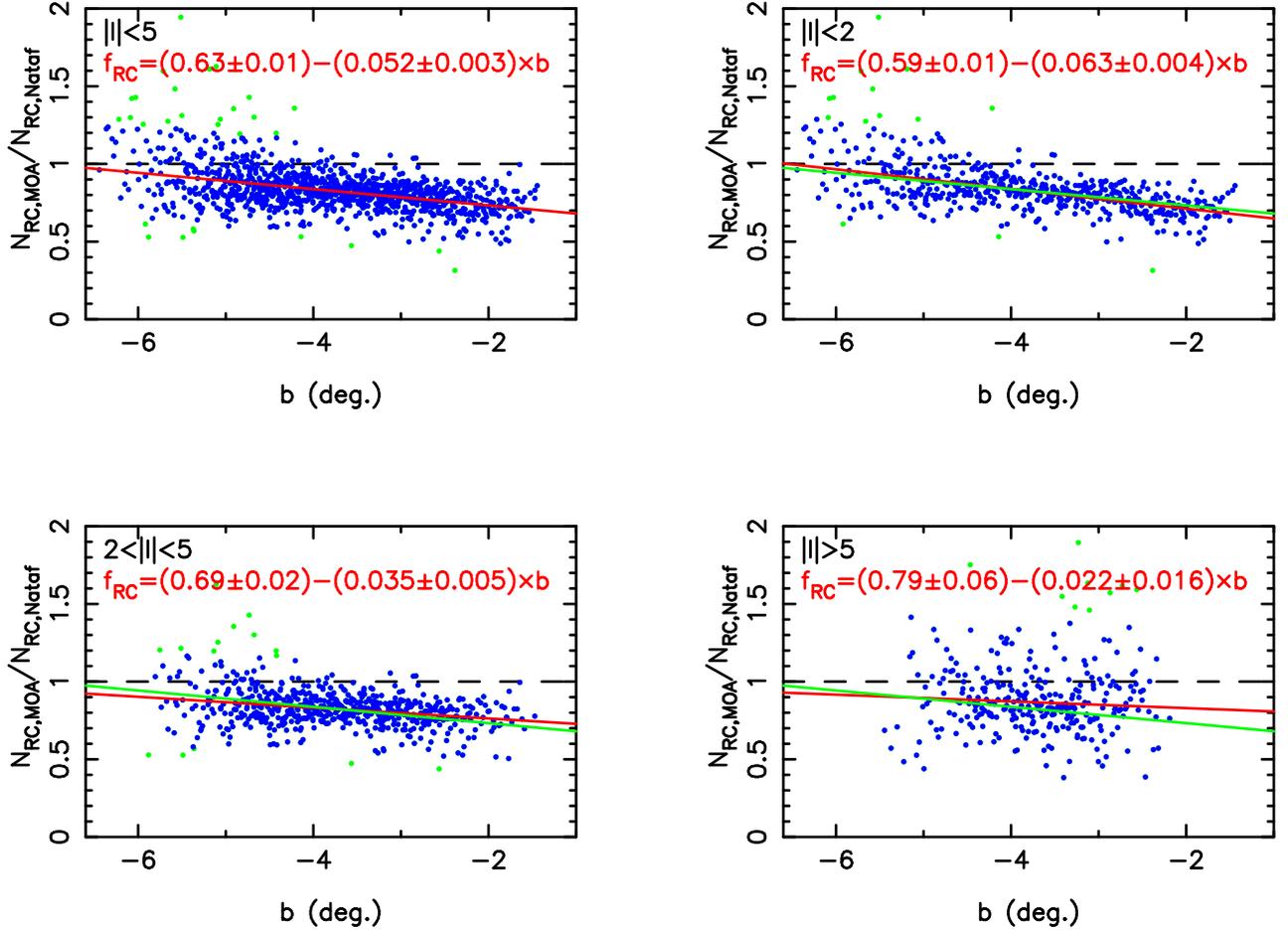}
\caption{
  \label{fig:fRC_b}
The number counts of RCG by MOA over that by OGLE, $f_{\rm RC}=N_{\rm RC,MOA}/N_{\rm RC,Nataf}$s for the subfields. The top-left, top-right, bottom-left and bottom-right panels are for subfields with
the galactic longitude of $|l| \le 5^\circ$,  $|l| \le 2^\circ$,  $2^\circ<|l| \le 5^\circ$ and  $|l|>5^\circ$,  respectively.
The red lines indicate the best fit using the blue dots, where 
3$\sigma$ outliers (green dots) are recursively rejected. 
The green lines indicate the best fit for $|l| \le 5^\circ$ as a comparison. They are all consistent with each other.
}
\end{center}
\end{figure}

In order to correct the number counts for incompleteness, we fit for relations between 
$f_{\rm RC}$ as a function of $N_{\rm RC,Nataf}$,  $I_{\rm RC,Nataf}$ and $b$, which are shown in Figure \ref{fig:fRC_NRC}, Figure \ref{fig:fRC_IRC} and the 
top-left panel of Figure \ref{fig:fRC_b}, respectively.
The scatter in these figures, as well as Figure \ref{fig:NRC_Nataf_MOA}, are about 10\%. 
In principle, if both data sets were equally complete, there should be no scatter between the two. 
The Poisson uncertainty on the number of stars that are missed in the MOA data is expected to be a few percent.
The variation of incompleteness in different subfields can generate additional scatter. 
We also expect that some scatter is caused by the averaging of Nataf et al.'s subfields in order to match a MOA subfield, 
together with the fact that the sky covered by the averaged Nataf et al. subfields is not exactly the same as the sky covered by the MOA subfield.
Here we perform the linear fits with recursive 3$\sigma$ clipping. In all figures, we can 
see clear trends. The standard deviations of the residuals for $N_{\rm RC,Nataf}$,  $I_{\rm RC,Nataf}$ and $b$, after (before) 3$\sigma$ clipping, 
are 0.10 (0.13), 0.11 (0.14) and 0.10 (0.13), respectively.
Thus the relation with $N_{\rm RC,Nataf}$ and $b$ are better than that of $I_{\rm RC,Nataf}$.
As the aim of the  this comparison is to correct the number counts for the completeness,
it could bias the result if we use $N_{\rm RC,Nataf}$ itself to correct for it, as they have their own uncertainty
as mentioned above.
Thus we decided to use the relation,
\begin{equation}
  \label{eq:fRC_b}
f_{\rm RC}=\frac{N_{\rm RC,MOA}}{ N_{\rm RC,Nataf} }=(0.63\pm0.01)-(0.052\pm0.003)\times b.
\end{equation} 

In Figure \ref{fig:fRC_b}, we also show $f_{\rm RC}$ as a function of $b$ with
the galactic longitude of  $|l| \le 2^\circ$,  $2^\circ<|l| \le 5^\circ$ and  $|l|>5^\circ$,  respectively.
One can see that this relation does not depend on the galactic longitude.
So, we apply the relation of Eq. (\ref{eq:fRC_b})  to all subfields.

\begin{figure}
\begin{center}
\includegraphics[angle=-90,scale=0.5,keepaspectratio]{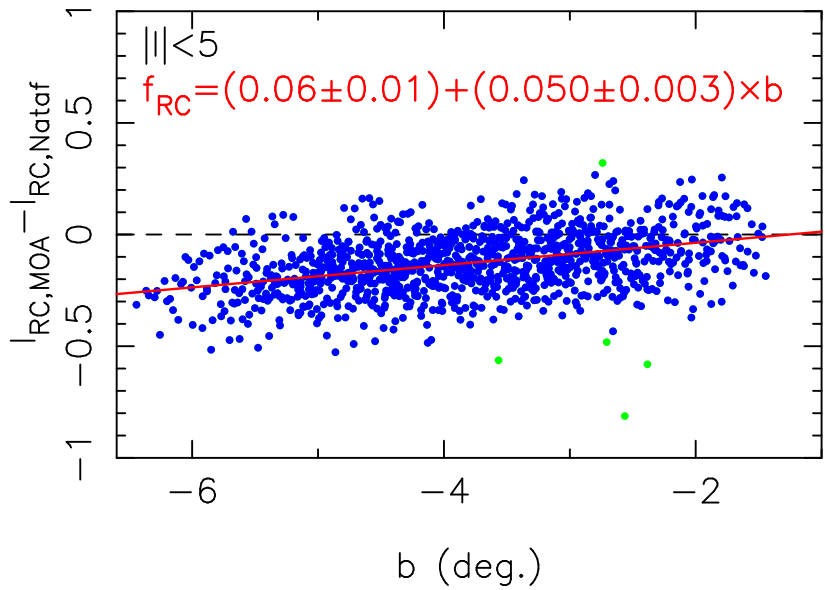}
\caption{
  \label{fig:dIRC_b}
The difference between $I_{\rm RC}$ by MOA and that by OGLE, $I_{\rm RC,MOA} - I_{\rm RC,Nataf}$
of subfields with the galactic longitude of $|l| \le 5^\circ$.
The red line indicates the best fit by using the blue dots where 
3$\sigma$ outliers (green dots) are recursively rejected. 
}
\end{center}
\end{figure}

If there is a significant systematic trend in  $I_{\rm RC,MOA}$, then it could also cause
a bias in the number count of sources for the all source sample. It is possible that the 
incompleteness of the RCG mentioned above might bias $I_{\rm RC,MOA}$.
We show the differences between $I_{\rm RC,MOA}$ and $I_{\rm RC,Nataf}$ as 
a function of $b$ in Figure \ref{fig:dIRC_b}.
One can see that there is a weak trend, while the amount of the difference due to
this slope over the range $b=-2\sim-6$  is comparable to the uncertainty in calibration between 
OGLE and MOA magnitudes which is about $\sim$0.2 mag.
This difference may also be due to the difference in the filter of MOA and OGLE but it is not clear at this stage.
The systematic trend in $I_{\rm RC}$ would affect 
estimates of the event detection efficiency in a complicated way in addition to 
affecting estimates of the number of sources. 
To see the magnitude of the bias due to systematics in $I_{\rm RC}$, 
we calculated the optical depth by correcting the correlation of 
$I_{\rm RC,MOA} - I_{\rm RC,Nataf} \propto 0.05\times b$.
We found that the difference 
from the optical depth results without this correction is a few percent or less than 10\% at a maximum.
Given the concordance of our updated optical depths 
with other data sets that would not be affected by this bias, we anticipate 
that the effect of the bias is smaller than our statistical uncertainties.

Because the reason for the trend in  $I_{\rm RC}$ is not clear and 
the effect is relatively small compared to the bias due to the incompleteness on the number count,
we  correct only for the effect of incompleteness on the number count and 
not for the effect on $I_{\rm RC}$ in the following analysis.

\subsection{Cause of the Incompleteness}

The \cite{Nataf2013} analysis is  conservative in order to ensure high completeness. 
To ensure the completeness in the $V$-band, they only consider fields where $(V-I)_{\rm RC} <=3.30$, 
which allows a maximum reddening of  $E(V-I)=2.24$ with the intrinsic RCG color 
of $(V-I)_{\rm RC,0}=1.06$ \citep{Bensby2013}. 
The RCGs are more complete in $I$-band, which are used for the number count, 
because they are 3.3 magnitude brighter than in $V$-band at this reddening.
The OGLE luminosity function of any field shows that the number counts drop off at $I\sim20.5$ mag, 
meaning completeness likely begins to fall at $I\sim 19$ mag or fainter.
In contrast, the maximum reddening of $E(V-I)=2.24$ implies $I_{\rm RC} = 17.1$ 
based on the RCGs at 8kpc with the intrinsic $I$-band magnitude of $M_{I,\rm RC,0}= -0.12$ and
the average total-to-selective extinction ratio of  $R_I = A_I/E(V-I)= 1.215$ \citep{Nataf2013}.
This is substantially brighter than the estimated completeness cutoff. 

For the RCG number count, \cite{Nataf2013} used stars with $V-I \ge (V-I)_{\rm RC} - 0.3$ and  $-1.5<I -I_{\rm RC}< 1.5$.
They included stars without $V$-band photometry assuming that they are fainter in $V$-band, i.e., redder than the limit $(V-I)_{\rm RC} - 0.3$. 
Thus their number count is not affected by the incompleteness in $V$-band.

On the other hand,
there some reasons for the incompleteness of RCG number count in MOA-II analysis by \cite{sumi2013}.
The major reason is the incompleteness in $V$-band catalog compiled using {\sc DoPhot}.
The limiting magnitude in $I$-band is $I\sim 18.3$ mag which is 
still deeper than the $I$-band RCG magnitude of $I_{\rm RC} = 17.1$ with the reddening of $E(V-I)=2.24$
given above, which is roughly equivalent to or slightly lower than the maximum reddening in MOA-II fields.
However, $V$-band limiting magnitude is about $V\sim 20.3$ mag, which is comparable 
to the $V$-band RCG magnitude of $V_{\rm RC} = 20.4$ with $E(V-I)=2.24$, i.e., 
$A_V= 4.96$ by using  $R_I = A_I/E(V-I)= 1.215$ \citep{Nataf2013}.

\cite{sumi2013} computed the RCG number count using stars selected with a similar color and magnitude limit as OGLE,  
$I < 17.5$ mag, $V-I \ge (V-I)_{\rm RC} - 0.3$ as shown in their Figure 1. However, stars without $V$-band 
photometry were not included. Thus the incompleteness in $V$-band catalog 
affects the number count.

We conclude that the RCG number count in MOA-II was not optimized for this purpose.
There are some ways to avoid this problem in future analysis for MOA-II. For example,
(1) get deeper $V$-band images, 
(2) Include all stars without a $V$-band detection following \cite{Nataf2013},
(3) Use the OGLE number count instead.

\section{Microlensing Optical Depth and Event Rate}
\label{sec:opt}

Here we have re-calculated the microlensing optical depth and event rate
by following  \cite{sumi2013}, but using the stellar number count 
corrected for the completeness by Eq.(\ref{eq:fRC_b}).

The optical depth, $\tau$, can be calculated by using Eq. (\ref{eq:opt}).
The microlensing event rate per star per year, $\Gamma$, can be determined observationally
from the following expression,
\begin{equation}
  \label{eq:Gamma}
  \Gamma = \frac{1}{N_{\rm s}T_{\rm o}} \sum_i \frac{1}{\varepsilon (t_{{\rm E},i})},
\end{equation}
Here we use the detection efficiency determined by \cite{sumi2011}.
 In our event rate and optical depth analyses for this 2006-2007 data set, 
 $T_{\rm o}=596.0$ days and the corrected number of source stars is 
 (1) $N_{\rm *}=110.3\times 10^6$ for the all-source sample and  
(2) $N_{\rm *,RC}=8.00\times 10^6$ for the RCG sample.


Individual optical depth estimates for all sources in each field are listed in Table~\ref{tbl:fld}.
The upper panel of Figure~\ref{fig:fieldmaps} shows a smoothed map of
optical depth of each subfield in Galactic coordinates. 
The plotted values from all subfields are listed in Table \ref{tbl:opt_2D} of
the online version, with a sample of this table listed in the printed version of this paper.
The smoothing is done with a Gaussian function with $\sigma=0.4^\circ$, and cut off
at a distance of $1^\circ$ from the center of each subfield.
The error bars for each subfield are estimated with the bootstrap method of \cite{alc97}, using the
neighboring subfields with the same weighting as in the calculation of the central values.

We also estimated the average optical depth in all fields combined, and found
$\tau_{200}= 1.53_{ -0.11}^{+  0.12}\times 10^{-6}$ with 474 events for all source sample and
$\tau_{200}= 1.28_{-0.19}^{+0.27}\times 10^{-6}$ with 83 events for RCG sample 
at $(l,b)=(1.^\circ85, -3.^\circ69)$. These are reductions of $18$ and $19$~percent, respectively, or $2.6\sigma$ and $1.3\sigma$, respectively.
The effective line of sight was 
computed by weighting the number of subfields used. The errors were 
estimated using the bootstrap Monte-Carlo method of \cite{alc97}.

\section{Modeling the Optical Depth Results}
\label{sec:model_opt}

The optical depth given by Equation~(\ref{eq:opt}) does not
follow Poisson statistics because each event is summed
with an unequal weight of $t_{{\rm E},i}/\varepsilon (t_{{\rm E},i})$.
Therefore, we binned the optical depth values of the subfields in order to model
the optical depth distribution.

Figure~\ref{fig:optball} shows the optical depth, $\tau_{200}$, as a function of $b$
for both the all-source sample and RCG samples for the central region with $|l|<5^\circ$, chosen so as to 
overlap with previous measurements.
The subfield results are binned with a bin width of $\Delta b= 0.5^{\circ}$.
The binned values for the all-source and RCG  samples are given in 
Tables \ref{tbl:opt_binb_all_lth5} and  \ref{tbl:opt_binb_RCG_lth5}, 
respectively.

The optical depth clearly increases with decreasing $|b|$, and a
simple exponential fit gives,
$\tau_{200}= [1.84\pm 0.14]\times10^{-6} \exp[(0.44\pm 0.07)(3-|b|)]$ for 
the all-source sample as indicated by the black solid line in Figure~\ref{fig:optball}. 
This is a significantly lower and shallower slope than the original result of
$\tau_{200}= [2.35\pm 0.18]\times10^{-6} \exp[(0.51\pm 0.07)(3-|b|)]$ estimated by \cite{sumi2013}
before correcting the completeness of the RCG number count.
The exponential model still represents the data well.

This result is significantly smaller than
the measurements by MOA-I \citep{sumi03} and MACHO \citep{alc00b} 
with all-source samples. Contrary to the original measurements, it is very consistent with 
the RCG measurements by MACHO \citep{pop05},
EROS-2 (\citealt{ham06}) and OGLE-II \citep{sumi2006}.
 The best linear fit to the OGLE-II RCG measurements is indicated by the red dashed line 
in Figure~\ref{fig:optball} as a comparison. 

 The MACHO \citep{pop05} and EROS \citep{ham06}
analyses identified microlensing events 
solely by their proximity to apparent RCG stars identified in the reference
images, with no attempt to determine if the source is a RCG star or a blended fainter
main sequence star.  These blending
effects will shrink apparent $t_{\rm E}$ values for all events, while increasing
the number of apparent RCG events. \citet{pop05} and \cite{ham06} 
make arguments to suggest that these two effects approximately cancel.

The only previous RCG sample that distinguished RCG source events from 
events with main sequence sources that happened to be blended with
RCG stars was the OGLE-II analysis of \citet{sumi2006}. I.e., they are
less affected by blending than the above MACHO and EROS RCG analyses.
The one similarity of their RCG analysis with other RCG analyses is that they 
require that events to be associated with resolved stars.
Their OGLE-II value is consistent with MACHO and EROS RCG analyses.
Thus it is likely that this cancellation of shrinking $t_{\rm E}$ and increasing 
the number of events, works to within the accuracy presented in their analyses,
as confirmed by image level simulation by \cite{smith2007}.

An exponential fit for the optical depth toward RCG sources gives
$\tau_{200}= [1.28\pm 0.21]\times10^{-6} \exp[(0.40\pm 0.17)(3-|b|)]$,
which is indicated by the red solid line in the Figure~\ref{fig:optball}.
This is also significantly lower than the original estimate and
previous RCG measurements
\citep{pop05,sumi2006,ham06} and some older bulge models 
\citep{bis02,han03,kerins2009}.

However, in this particular analysis, this $\tau$ estimate for the RCG sample is heavily biased by the low detection efficiency for
events with $t_{\rm  E} > 100\,$days, due to the fact that the analysis was
originally designed to focus on short time scale events. 
Although the events with $t_{\rm E}<200$days are selected, most of the events with $t_{\rm  E} > 100\,$days
with bright sources could not satisfy the requirement for a long enough constant baseline.
This is because the tails of the events are longer than $t_{\rm E}$ and they are 
still significantly above the baseline for bright source events.
This effect is negligible for the all source sample.
Thus we can not directly compare this result with other measurements.

\section{Modeling the Event Rate}
\label{sec:model_gamma}

The event rate per square degree per year, $\Gamma_{\rm deg^2}$,
for source stars above a magnitude threshold of $I_s \leq 20$,
which are given in Table \ref{tbl:fld}-\ref{tbl:opt_2D} for completeness,
does not change from \cite{sumi2013} because this quantity is independent 
of the stellar number count. Thus, we focus on 
the event rate per star per year, $\Gamma$, in the rest of the paper.

We model the event rate, $\Gamma$ by using the Poisson statistics fitting method,
first introduced by \cite{sumi2013}.  This method allows us to fit to the raw, subfield data, 
even though the average number of events per subfield is $<1$, thus 
free from the problem on the binning of the sample and the improper assumption of the gaussian statistics.
The number of expected events in a subfield is given by
\begin{equation}
  \label{eq:Nevexp}
  N_{\rm ev, exp}(l,b)=\Gamma_{\rm mod}(l,b) N_{\rm s}(l,b)T_{\rm o} \langle\varepsilon(l,b)\rangle \ ,
\end{equation}
where $N_{\rm s}$ is the number of stars in the subfield,
and $\langle\varepsilon(l,b)\rangle$ is the detection efficiency averaged over $t_{\rm E}$ for
the subfield at coordinates $(l,b)$. 
We adopt the average detection efficiency given in Tables 2 and 3 of \cite{sumi2013}, for the
all-star and RCG samples, respectively, while  $N_{\rm s}$ need to be corrected by Eq. (\ref{eq:fRC_b}).

The probability of the observed number of events, $N_{\rm ev}(l,b)$, in the subfield at $(l,b)$ is
\begin{equation}
  \label{eq:Poisson}
  P[N_{\rm ev}(l,b)]= \frac{ e^{-N_{\rm ev, exp}(l,b)} N_{\rm ev, exp}(l,b)^{N_{\rm ev}(l,b)}}{N_{\rm ev}(l,b)!} \ ,
\end{equation}
according to Poisson statistics. We can then define the $\chi^2$ by,
$
  \chi^2 = -2 \sum_{(l,b)} \ln P[N_{\rm ev}(l,b)].
$

Thus the event rate, $\Gamma$, is the preferred
quantity to compare to Galactic models rather than the optical depth.
We show the event rate per star per year $\Gamma$ and the exponential fits for the 
all-source and RCG samples as a function of the galactic latitude, $b$, for $|l|<5^\circ$
in Figure \ref{fig:Gamma_vs_b} and in Tables~\ref{tbl:opt_binb_all_lth5}
and \ref{tbl:opt_binb_RCG_lth5}, respectively.  
The event rate has much less scatter
than $\tau_{200}$ and $\Gamma$ for both the all-source and RCG samples, and both
are well fit by a simple exponential model. Note that these fits are done to the
subfield data using the Poisson statistics method, 
while the plots show binned quantities for display in Figure \ref{fig:Gamma_vs_b} and in 
Tables~\ref{tbl:opt_binb_all_lth5} and \ref{tbl:opt_binb_RCG_lth5}.

The exponential model for the all-source and RCG samples are quite
similar with
$\Gamma_{\rm all}= [18.74\pm 0.91]\times10^{-6} \exp[(0.53\pm 0.05)(3-|b|)] \,{\rm star}^{-1}{\rm yr}^{-1}$ 
for the all-source sample and
$\Gamma_{\rm RC}= [17.13\pm 2.03]\times10^{-6} \exp[(0.58\pm 0.12)(3-|b|)] \,{\rm star}^{-1}{\rm yr}^{-1}$
for the RCG sample. 
Again, due to the $N_{\rm s}$ correction, these have a smaller and shallower slope than the original values in \cite{sumi2013}.
The RCG event rate is slightly smaller, 
but consistent with the all-source
event rate. The RCG slope is $0.4\sigma$ steeper and the amplitude is
is 8\% or $0.8\sigma$ smaller. As noted earlier, this is because 
$\Gamma$ is much less sensitive to the bias due to the small number of long $t_{\rm E}$ events.


\cite{sumi2013} noted that although there is a possible problem with assuming the luminosity function
in all other fields are same as the {\it HST} luminosity function measured
in Baade's window \citep{hol98}, 
the uncertainty due to the different luminosity function shape would
largely cancel out if the same luminosity function is used
in the detection efficiency simulations and the source star counts.
The consistency between the all-source and RCG $\Gamma$ values
indicates that the effect due to the variation of the luminosity function shape in
each field relative to the {\it HST} luminosity function are negligible.
However, they were not aware that the effect on the normalization of the luminosity function is more significant than its shape.

We show exponential fits as a function of the galactic latitude $b$ for 
$\tau_{200}$ and $\Gamma$  for different bins 
in Galactic longitude, $l$ in Figures
\ref{fig:tau_vs_b_l} and
\ref{fig:Gamma_vs_b_l}, respectively.
The black points and curves are for all the events with
$-2^\circ.25 < l < 3^\circ.75$.
In  Figure \ref{fig:Gamma_vs_b_l}, it provides a reasonable fit to all the
longitude bins, except the $0^\circ.75 < l < 2^\circ.25$ bin,
where there is an enhancement to the rate.
On the other hand, there is some scatter in $\tau_{200}$ between different bins in Figure \ref{fig:tau_vs_b_l}.
The $\tau_{200}$ bin with $0^\circ.75 < l < 2^\circ.25$  at small $|b|$ is 
smaller than the average, which is different from $\Gamma$.
This is because average $t_{\rm E}$ is smaller at these galactic central regions 
due to the galactic kinematics as shown in Fig. 3 of   \citet{sumi2013}.

Figure \ref{fig:fieldmaps} shows smoothed maps of
$\tau_{200}$ and $\Gamma$ in Galactic
coordinates. The plotted values from all subfields are listed in Table \ref{tbl:opt_2D}.
The smoothing is done with a Gaussian function with $\sigma=0.4^\circ$, and cut off
at a distance of $1^\circ$ from the center of each subfield.
The error bars for each subfield are estimated using a bootstrap method using the
neighboring subfields with the same weighting as in the calculation of the central values.
They are similar to the original maps in Fig. 3 of  \cite{sumi2013}, but 
decreased by up to $\sim40$\% depending on $b$.
The highest optical depth is found at $l\approx3^\circ$ and this is due to the excess of 
long timescale events at this longitude, and could be due to the statistical 
fluctuations enhanced by large weight for long events.


The event rate per star, $\Gamma$ has a peak at $l\approx 1^\circ$. 
Because these event rate measurements obey
Poisson statistics, the statistical uncertainty in $\Gamma$ 
is smaller than the uncertainty in $\tau_{200}$. 
So, we expect that this $l\approx 1^\circ$
enhancement in the microlensing rate is real and that it is related to the structure
and kinematics of the bulge.

As a comparison to \cite{sumi2013}, we have fit $\Gamma$ with a 16-parameter model in $l$ and $b$. The
16 parameters consist of a 10-parameter cubic polynomial and the inverse of
a 6-parameter quadratic polynomial. That is
\begin{equation}
  \label{eq:2Dformula}
\begin{eqalign}
\Gamma = \ &
a_0  + 
a_1 l +
a_2 b+ 
a_3 l^2+
a_4 lb+
a_5 b^2+
a_6 l^3+
a_7 l^2b+
a_8 lb^2+
a_9  b^3 \\
& + 1/(
a_{10}  + 
a_{11} l +
a_{12} b+ 
a_{13} l^2+
a_{14} lb+
a_{15} b^2
) \ .
\end{eqalign}
\end{equation}
The best fit model is shown in Figure \ref{fig:Gamma_2D}  
and the model parameters are listed in Table~\ref{tbl:param_2D}.
The model has a maximum at $l\approx 1^\circ$ that was also
evident in Figure \ref{fig:fieldmaps}.

\section{Discussion and conclusions}
\label{sec:discussionAndSummary}

\label{sec:summary}

We examined the completeness of the stellar number count in 
the measurement of the microlensing optical depth $\tau$ and event rate per star per year $\Gamma$ toward the 
Galactic bulge from the first two years of the MOA-II survey \citep{sumi2013}. 
We found a significant incompleteness in MOA-II's RCG counts,
which is proportional to the galactic $b$. The completeness rangs from 1 to 0.7 for $b=-6^\circ\sim-1.5^\circ$.
The counts are less complete at lower $|b|$ because of the higher stellar number density and the
higher interstellar extinction. This incompleteness caused the overestimates in the $\tau$ and $\Gamma$.

By correcting this incompleteness, we estimated $\tau$ and $\Gamma$ 
with the all source sample of 474 events and a RCG sample of 83 events.
Note that our RCG optical depth is known to be biased low, due to the low efficiency
for long duration bright events. Thus we focus on $\tau$ with all source sample in the following discussion. 
Due to this correction of the incompleteness, both $\tau$ and $\Gamma$ decreased at lower $|b|$.
This result may have solved the previously
noted difference between the optical depths measured with RCG samples
\citep{pop05,sumi2006,ham06} and that with faint source samples from DIA
\citep{alc00b,sumi03}, for which the faint source analyses
have shown systematically higher $\tau$ values.

The original measurement of $\tau$ with all source by MOA-II \citep{sumi2013} were in-between of the other 
previous measurements with all-source and RCG sample and consistent within 1 or 2-$\sigma$ level. 
Thus they concluded that the previously seen difference between the all-source and RCG samples was
due to statistical fluctuations.


However, our revised all-source optical depth measurements are consistent with
previous measurements for RCG samples and significantly 
lower than that of the \citet{sumi2013} all-source sample  as shown in Figure \ref{fig:optball}.
We can use the exponential models shown in Figure \ref{fig:optball} to interpolate
our measurement to the center of previous samples. For the MACHO DIA 
all-source result at $b = -3^\circ .35$ \citep{alc00b}, we find 
$\tau_{200} = [1.58\pm 0.13] \times 10^{-6}$ which
is 2.1-$\sigma$ smaller than the MACHO result of 
$\tau_{180}  = 2.43{+0.39\atop -0.38}\times 10^{-6}$. 
The MOA-I all-source result \citep{sumi03}, centered at 
$b = -3^\circ .8$, is $\tau_{150} = 2.59{+0.84\atop -0.64}\times 10^{-6}$. This compares to
our interpolated value of $\tau_{200} = [1.29\pm 0.11]\times 10^{-6}$, which is 1.7-$\sigma$ smaller.
Thus, our revised optical depth measurement is smaller at the $\sim$\,2-$\sigma$ level than both 
the previous all-source measurements, which suggests that 
these measurements also suffer the same bias in
the stellar number count. Because \cite{alc00b,sumi03} used the similar method as this 
work, it is very likely that they have similar bias.

The MACHO
Collaboration published several averages of their results \citep{pop05}, but
we compare to their ``CGR+3" average of $6\,{\rm deg}^2$ centered at
$b = -2^\circ.73$. MACHO reports $\tau_{500} = 2.37{+0.47\atop -0.39}\times 10^{-6}$
for RCG sources at this position. This compares to our interpolated 
all-source value of $\tau_{200} = [2.08\pm 0.17] \times 10^{-6}$, which is
just 0.6-$\sigma$ smaller. 

The RCG sample of the EROS Collaboration \citep{ham06} covers a
slightly larger area than the MOA-II analysis.
They fit their results to an exponential model that is identical to the one
shown in Figure \ref{fig:optball}, and they find
$\tau_{400} = [1.62\pm 0.23]\times10^{-6} \exp[(0.43\pm 0.16)(3-|b|)]$. This has a
slope that is consistent with our fits,
so we compare the results by simply comparing the
normalization parameters. Our normalization parameter is $[1.84\pm 0.14 ]\times10^{-6} $, which is 
0.8-$\sigma$ larger than the EROS value of $[1.62\pm 0.23 ]\times10^{-6}$.
A more fair comparison would be to compare to the the EROS
fit to a model fit to all our fields, instead of just those with $|l| < 5^\circ$.
This gives $\tau_{200} = [1.74\pm 0.13]\times10^{-6} \exp[(0.45\pm 0.07)(3-|b|)]$,
which is 0.4-$\sigma$ larger than the EROS value.

The OGLE-II RCG analysis \citep{sumi2006} 
 found $\tau_{400}  = 2.55{+0.57\atop -0.46}\times 10^{-6}$ at $b = -2^\circ .75$
which is consistent with other RCG analyses.
This compares to our all-source result, interpolated from the model given in 
Figure \ref{fig:optball}, is $\tau_{200} = [2.06\pm 0.17] \times 10^{-6}$, which is
just 0.9-$\sigma$ smaller.

 
In summary, we find that our all-source results are about 2-$\sigma$ smaller
than the previous all-source measurements, and they are very consistent
with the RCG optical depth values from OGLE, MACHO and EROS within 1-$\sigma$. 

Note that the observed optical depths mentioned above represent the contributions 
of optical depth from the events 
within the given timescale range. The upper limit of the $t_{\rm E}$ in these analyses 
range between 150-500 days.
However, the longest $t_{\rm E}$ detected in their sample is mostly less than 200 days and
the contributions of $\tau$ from the events with $t_{\rm E}>150$ days are negligible. 
Thus above comparison is valid within the their uncertainty.

We compare the optical depth results with values predicted from the models.
\citet{han03} model predicts  $\tau = [1.63\pm 0.13]\times10^{-6}$  at $b = -3^\circ.9$,
where our all-source optical depth $\tau_{200} = [1.24\pm 0.10]\times10^{-6}$ is 
3.9-$\sigma$ smaller.
The values from \cite{wood05} models are 
$\tau = 2.1\times10^{-6}$  at $b = -3^\circ$ and
$\tau = 3.0\times10^{-6}$  at $b = -2^\circ$.
Our values
$\tau_{200}  = [1.84\pm 0.15]\times10^{-6}$ and
$\tau_{200}  = [2.87\pm 0.24]\times10^{-6}$ agree with them
with 1.7-$\sigma$  and 0.5-$\sigma$, respectively.
\citet{eva02} present a number of models, and the value of their
``Dwek plus spiral structure" model
$\tau = 1.5\times10^{-6}$  at $b = -3.8^\circ$
agrees with our 
$\tau_{200}  = [1.29\pm 0.11]\times10^{-6}$ which is
1.9-$\sigma$ smaller, while their other 
models predict much higher optical depths. 
The models of \citet{bis02} predict  
$\tau = 1.1\times10^{-6}$  at $b = -3.35^\circ$ for all sources and 
$\tau = 1.3\times10^{-6}$  at $b = -3.8^\circ$  for RCG sources.
Our values 
$\tau_{200}  = [1.58\pm 0.13]\times10^{-6}$ and
$\tau_{200}  = [1.29\pm 0.11]\times10^{-6}$ are 3.7-$\sigma$  and 0.2-$\sigma$ larger, respectively.
\citet{kerins2009}'s model predicts
 $\tau = 4\times 10^{-6}$ at $b = -1^\circ.9$ and
 $\tau = 2\times 10^{-6}$ at $b = -3^\circ.5$, where
our measurements are
$\tau_{200}  = [3.00\pm 0.25]\times 10^{-6}$ and
$\tau_{200}  = [1.48\pm 0.12]\times 10^{-6}$
 which are 4.0-$\sigma$ and 4.3-$\sigma$ smaller, respectively. 

\begin{figure}
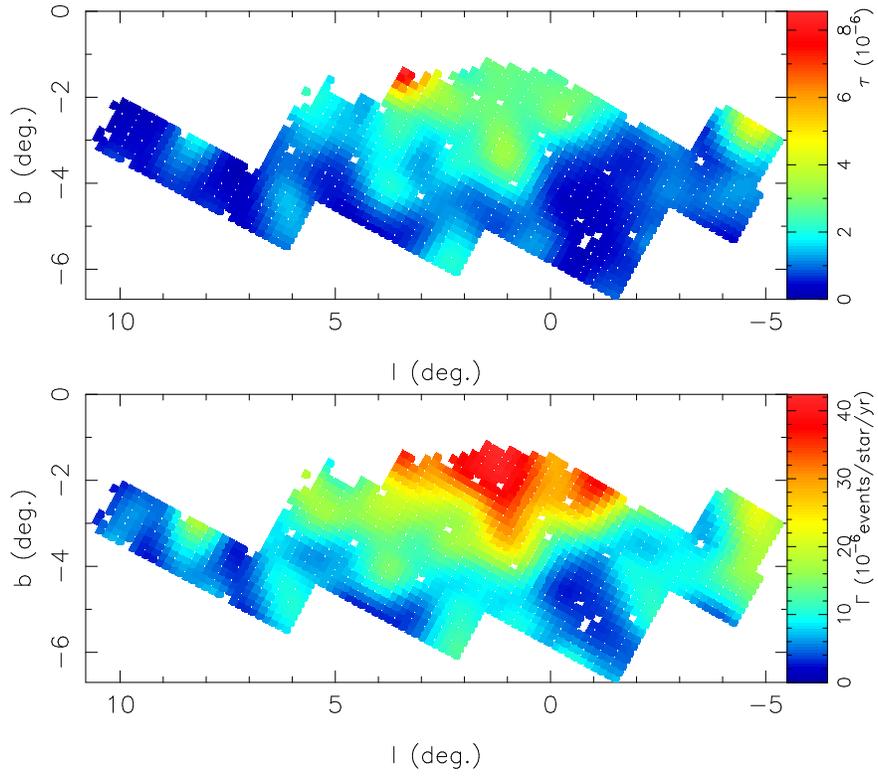

\begin{center}
\includegraphics[angle=-90,scale=0.45,keepaspectratio]{Figure8a.eps}
\includegraphics[angle=-90,scale=0.45,keepaspectratio]{Figure8b.eps}
\caption{
  \label{fig:fieldmaps}
False color maps of the measured optical depth, $\tau_{200}$ (top panel) and
the event rate per star per year, $\Gamma$ (bottom panel).
}
\end{center}
\end{figure}

\begin{figure*}
\begin{center}
\includegraphics[angle=-90,scale=0.65,keepaspectratio]{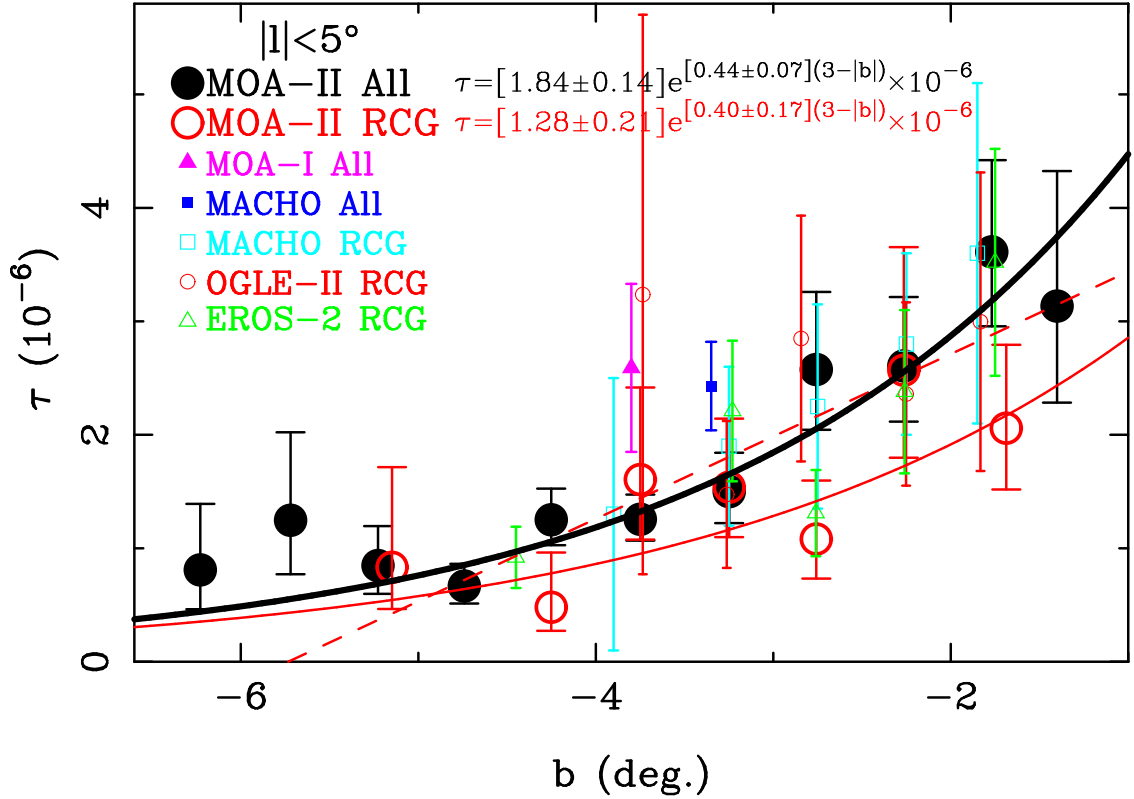}
\caption{The measured optical depth for the all-source (black filled circle) and 
RCG (red large open circle) samples as a function of galactic latitude $b$ for $|l|<5^\circ$. 
The subfields are combined into bins of width $\Delta b= 0.5^{\circ}$. The binned values are 
listed in Table \ref{tbl:opt_binb_all_lth5} and \ref{tbl:opt_binb_RCG_lth5}.
The filled circles, triangles and squares indicate $\tau$
for all-source samples measured by MOA-II (this work), MOA-I and MACHO surveys, respectively.
The red circles, open squares, circles and triangles denote the $\tau$ for RCG samples by the
MOA-II (this work), MACHO, OGLE-II and EROS surveys, respectively.
The thick black and thin red solid lines indicate the best fit exponential 
functions for the MOA-II measurements. The red dashed line denotes the best linear model
for the OGLE-II RCG sample as a comparison.
\label{fig:optball} 
}
\end{center}
\end{figure*}

\begin{figure*}
\begin{center}
\includegraphics[angle=-90,scale=0.65,keepaspectratio]{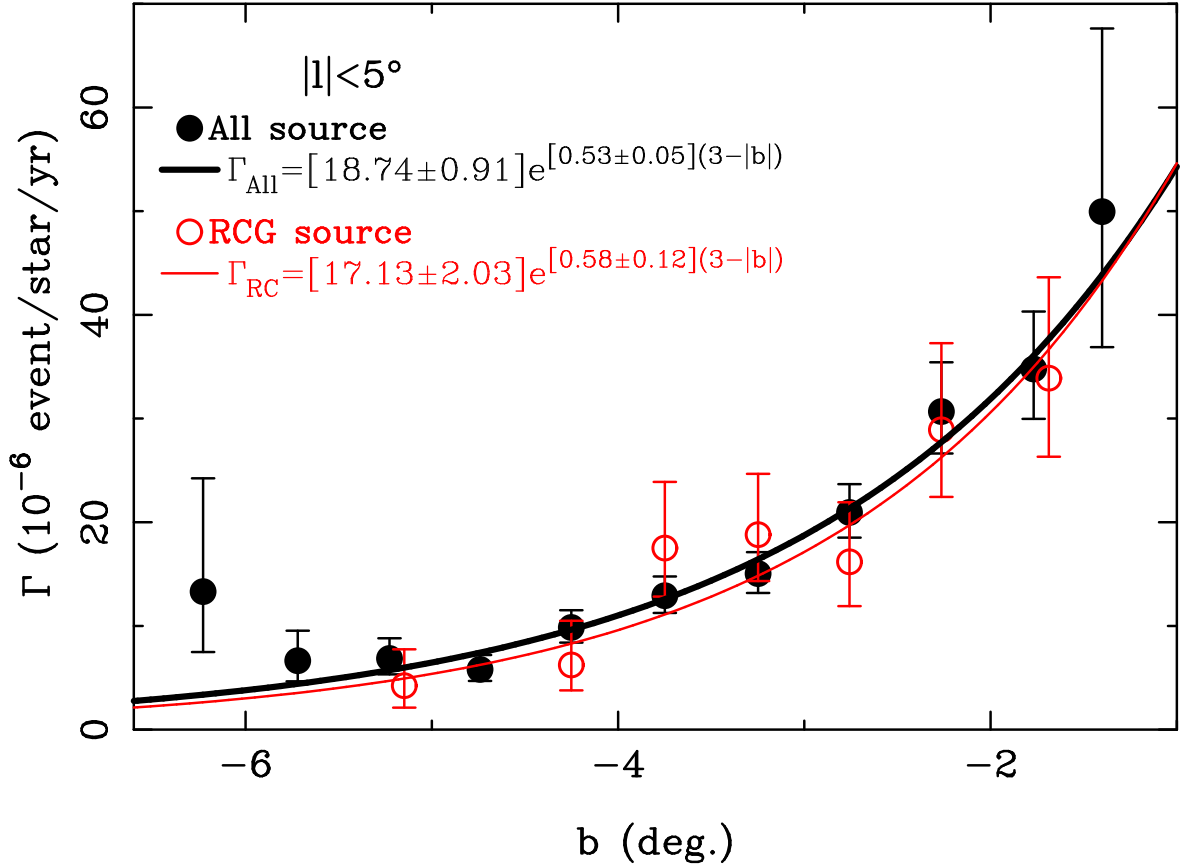}
\caption{The event rate per star per year, $\Gamma$, for the all-source (black filled circle) and RCG (red open circle) samples 
as a function of the galactic latitude $b$  for $|l|<5^\circ$. 
The subfields are combined into bins of width $\Delta b= 0.5^{\circ}$ for display purposes
only, as the fitting was done using the unbinned subfield data with the Poisson statistics
fitting method.
The plotted values are listed in Tables \ref{tbl:opt_binb_all_lth5}
and \ref{tbl:opt_binb_RCG_lth5}.
The thick black and thin red solid lines indicate the best fit exponential functions 
for the all-source and RCG samples, respectively.
\label{fig:Gamma_vs_b} 
}
\end{center}
\end{figure*}

\begin{figure*}
\begin{center}
\includegraphics[angle=-90,scale=0.65,keepaspectratio]{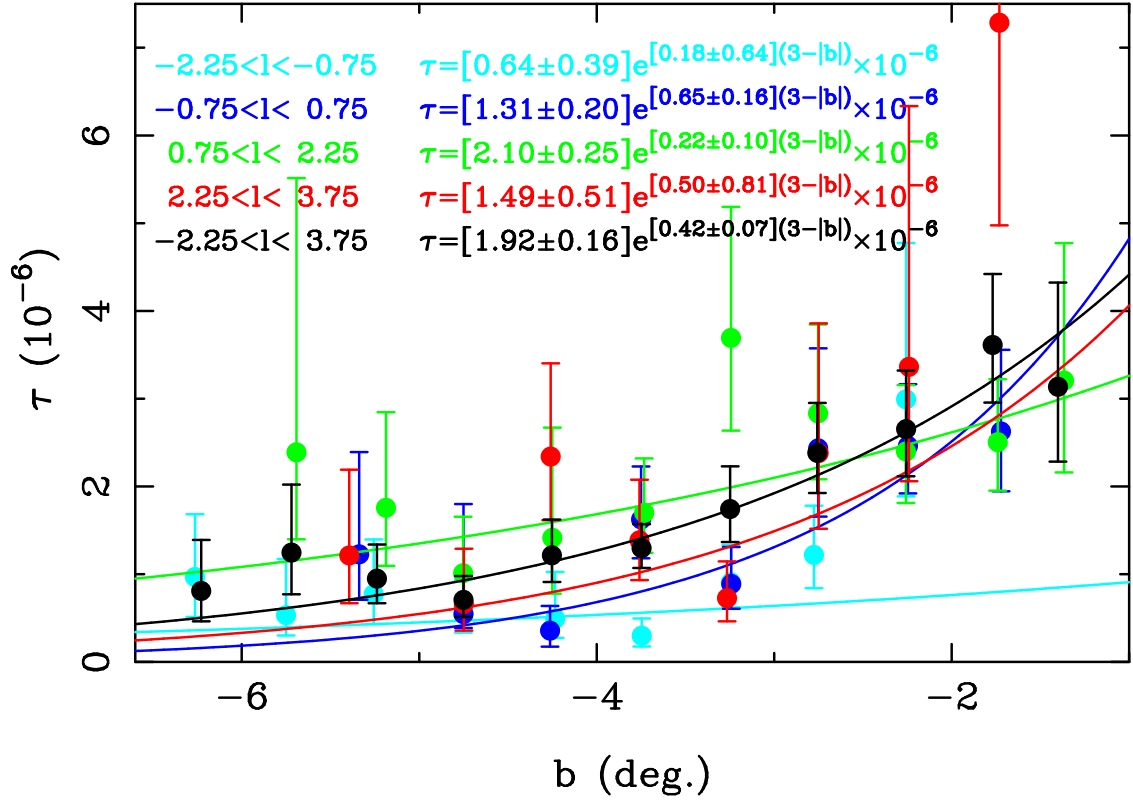}
\caption{The optical depth for events with $t_{\rm E} < 200\,$days, 
$\tau_{200}$, for the all-source sample 
as a function of the galactic latitude $b$ for different bins in Galactic
longitude, $l$. The curves show the best exponential fit in $b$.
The black curve is the fit to all the events with
$-2^\circ.25 < l < 3^\circ.75$. 
\label{fig:tau_vs_b_l} 
}
\end{center}
\end{figure*}

As discussed above in Section~\ref{sec:model_gamma}
and shown in Figures \ref{fig:fieldmaps} and \ref{fig:Gamma_vs_b}, $\Gamma$ 
can be measured more precisely than $\tau$. Furthermore, $\tau$ has an additional
systematic uncertainty due to potential very long time scale events, which may
contribute significantly to $\tau$ but not to $\Gamma$.
Figure \ref{fig:Gamma_vs_b}  indicates that the all-source and RCG $\Gamma$ values
differ by only less than $9$\%.

Recently, \cite{Awiphan2016}  presented the field-by-field comparison between 
results by \cite{sumi2013} and the Besan\c{c}on population synthesis Galactic model.
They found  only $\sim$50\% of the measured $\tau$ and event rate per star per year, 
$\Gamma$, at low Galactic latitude around the inner bulge ($|b| < 3^\circ$) and suggested 
the discrepancy most likely is associated with known underestimated extinction 
and star counts in the innermost regions, which supports missing inner stellar population.
Here we compared their model and our revised $\tau$ and $\Gamma$, in Fig. 
\ref{fig:awiphan_comp_tau} and \ref{fig:awiphan_comp_gam}, respectively.
Their model is more consistent with our revised $\tau$ than the original measurements by \cite{sumi2013}, but are still slightly higher.
The difference is not 
very significant due to the large error in our measurement.
Our revised $\Gamma$ are very consistent to their model
without any missing inner stellar population.

\cite{sumi2013} noted a possible problem with assuming that the luminosity function
in all fields are same as the {\it HST} luminosity function in Baade's window \citep{hol98}.  
However, the consistency between the all-source and RCG $\Gamma$ values 
indicates that the effect due to the variation of the luminosity function shape in
each field from the {\it HST}  luminosity function are negligible.

However, \citet{sumi2013} were not aware that the completeness of the number counts of RCGs might be problematic,  
because they were thought to be bright enough to be complete. 
In their analyses for both all source sample and RCG sample, the stellar number count
is based on the stellar catalog in the reference images reduced by DoPHOT  \citep{sch93} algorithm,
where the combined ground+HST luminosity function are normalized by the RCG number counts
and the events were selected regardless of whether they are associated with resolved stars. 
Thus, both samples are biased by the same amount. This incompleteness is mostly because MOA-II data 
was taken under relatively poor seeing of $\sim1.8$ arcsec even in the reference images.
The measurements by MOA-I \citep{sumi03} and  MACHO \citep{alc00b}  also used a similar method with similar seeing, 
and are therefore expected to suffer the same problem.

Other measurements using RCG samples by  OGLE-II  \citep{sumi2006}, MACHO  \citep{pop05} and EROS  \citep{ham06} used a different method,
which selected only the events at the position of the resolved stars in the reference image by DoPHOT or similar algorithm.
Thus, the incompleteness affects both the number of events and stellar number count, thus they cancel each other out.

In conclusion, we believe the long-standing problem of the discrepancy between the high optical depth in all source sample 
by DIA and low optical depth with RCG sample can be explained by the incompleteness of the 
stellar number count.

These measurements of $\Gamma$ and $\tau$ have the highest spatial resolution so far 
thanks to our samples being the largest studied so far.
Our goal is to measure $\Gamma$ and $\tau$ precisely around the galactic bulge 
to constrain the barred Galactic bulge model. Currently MOA-II and OGLE-IV 
detect about 700 and 2,000 events a year, respectively. In the near future,
this analysis will be expanded by thousands of events.

\begin{figure*}
\begin{center}
\includegraphics[angle=-90,scale=0.65,keepaspectratio]{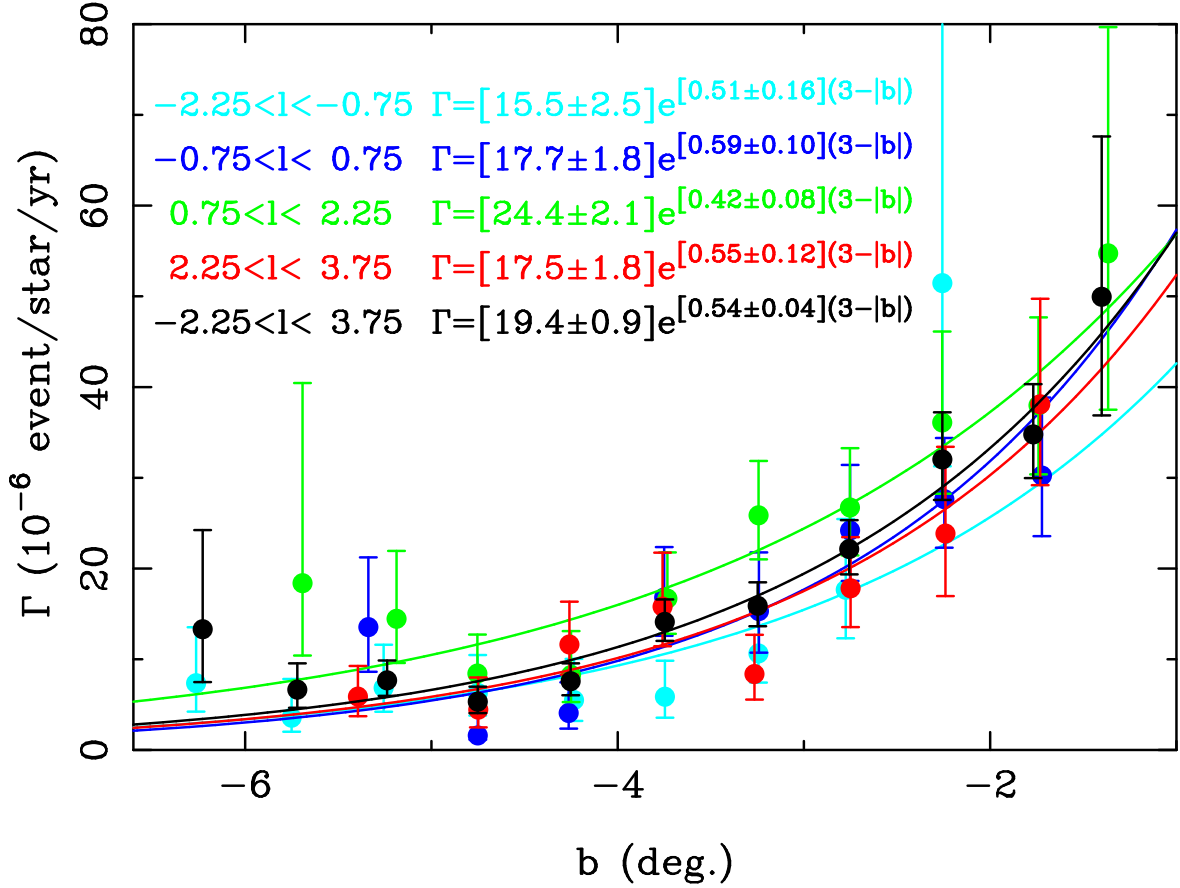}
\caption{The event rate per star per year, $\Gamma$, for the all-source sample 
as a function of the galactic latitude $b$ for different bins in Galactic
longitude, $l$. The curves show the best exponential fit in $b$ to the
unbinned subfield data. The black curve is the fit to all the events with
$-2^\circ.25 < l < 3^\circ.75$, and it provides a reasonable fit to all the
longitude bins, except the $0^\circ.75 < l < 2^\circ.25$ bin,
where there is an enhancement to the rate.
\label{fig:Gamma_vs_b_l} 
}
\end{center}
\end{figure*}

\begin{figure}
\begin{center}
\includegraphics[angle=-90,scale=0.6,keepaspectratio]{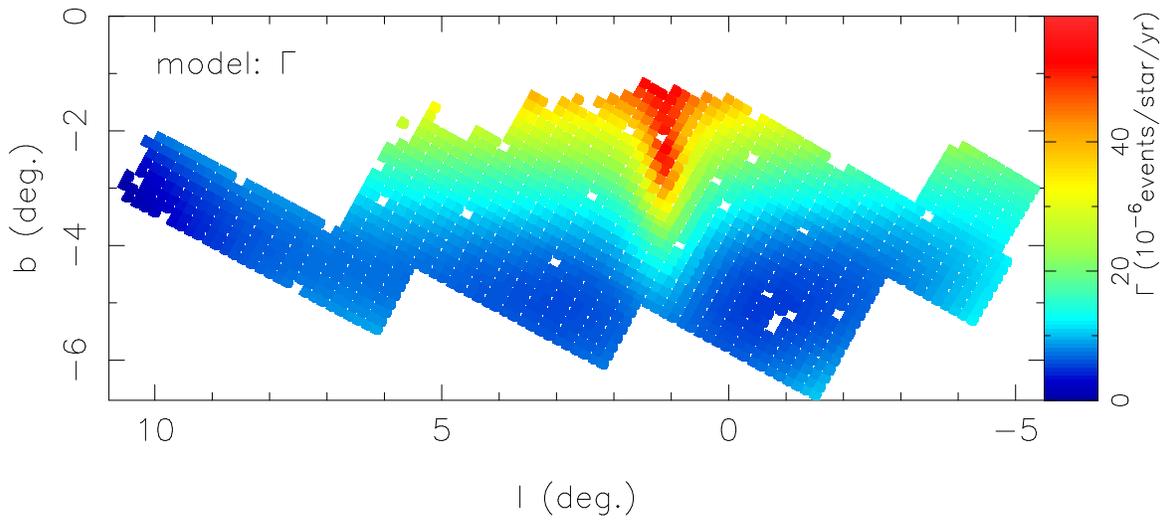}
\caption{
  \label{fig:Gamma_2D}
A 16-parameter model of microlensing event rate per star for the all-source sample. The 
model is described by Equation (\ref{eq:2Dformula}) with parameters given in Table~\ref{tbl:param_2D}.
}
\end{center}
\end{figure}

\begin{figure}
\begin{center}
\includegraphics[angle=90,scale=0.5,keepaspectratio]{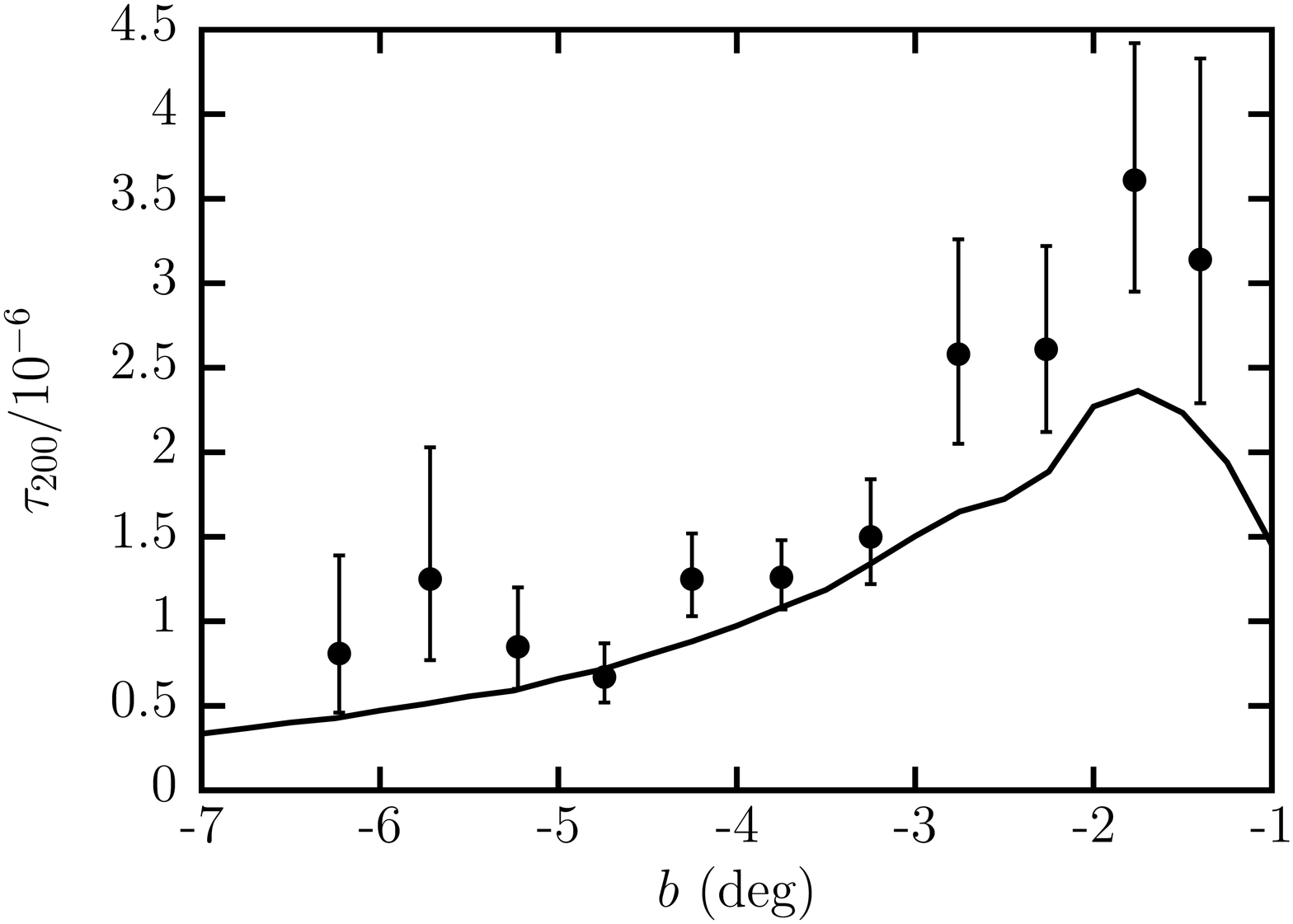}
\caption{
  \label{fig:awiphan_comp_tau}
  The optical depth for events with $t_{\rm E} < 200\,$days, 
$\tau_{200}$, for the all-source sample 
as a function of the galactic latitude $b$ (filled circles with error bars),
and the theoretical model from the Besan\c{c}on model by  \cite{Awiphan2016} (solid line).
It is better agreement than the original $\tau$ measurements by \cite{sumi2013},
while they are still slightly higher.
}
\end{center}
\end{figure}

\begin{figure}
\begin{center}
\includegraphics[angle=90,scale=0.5,keepaspectratio]{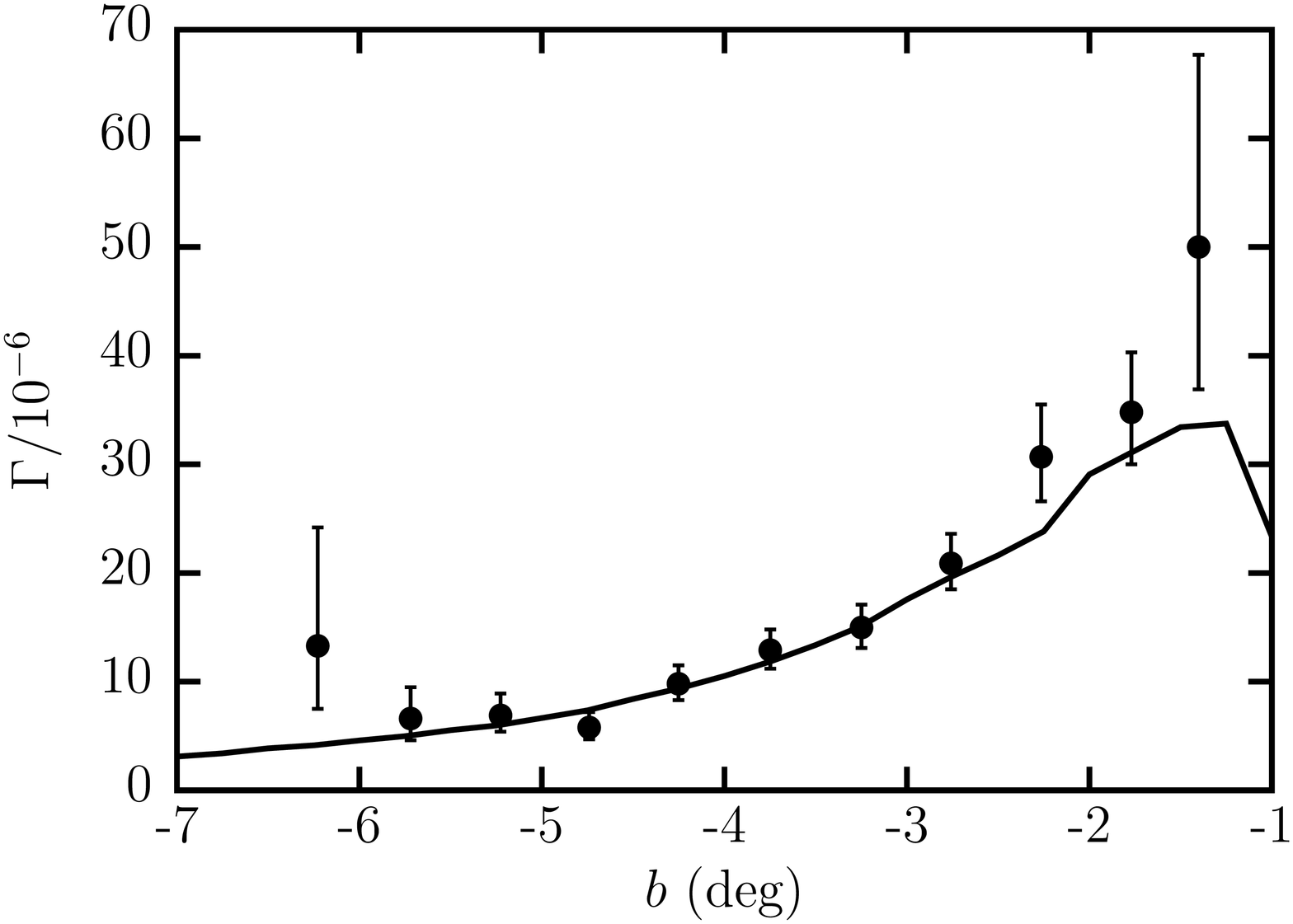}
\caption{
  \label{fig:awiphan_comp_gam}
The event rate per star per year, $\Gamma$, for the all-source sample 
as a function of the galactic latitude $b$ (filled circles with error bars),
and the theoretical model from the Besan\c{c}on model by  \cite{Awiphan2016} (solid line).
They are consistent.
}
\end{center}
\end{figure}

Another goal of this work is to predict the event rate in the inner 
Galactic bulge for the future space microlensing survey of the 
Wide Field Infrared Space Telescope (WFIRST) \citep{green2012,Spergel2015} and Euclid \citep{penny13}.
The expected microlensing event rate for the WFIRST mission is uncertain because the region
with the highest event rate at low Galactic latitudes are not well studied due to 
the high interstellar extinction.
This work provides the best estimate of the
event rate in the inner Galactic bulge to date. For $3.2\,{\rm deg}^2$ of
the MOA-II survey area inside $|b| \leq 3^\circ.0$ and $0^\circ.0 \leq l \leq 2^\circ.0$,
centered at  $(l,b)=(0.^\circ97, -2.^\circ26)$, we find 
$\Gamma =  3.41_{-0.34}^{+0.38}  \times 10^{-5}$\,star$^{-1}$\,{yr}$^{-1}$ for sources
with $I < 20$. This is consistent with the rate model used for the 
report of the WFIRST Science Definition Team (SDT) \citep{green2012,Spergel2015}
evaluated at this position, while the previous value was a factor 1.3 larger than this.
By extrapolating to the lower latitude fields, $|b| \sim -1^\circ.4$, where the WFIRST will 
observe, the model with $-2^\circ.25<l< 3^\circ.75$ presented in Figure~\ref{fig:Gamma_vs_b_l} predicts
$\Gamma =  4.60\pm 0.25  \times 10^{-5}$\,star$^{-1}$\,{yr}$^{-1}$  for sources with $I < 20$.
which is consistent with the value in the WFIRST SDT report.

In this work, we have attempted to correct for incompleteness by assuming that 
another data set is complete, when that data set itself has not been corrected 
for completeness~\citep{Nataf2013}. While we expect OGLE number counts to be more 
complete than MOA's, it was long assumed that completeness would not be an 
issue for MOA. This is of course not ideal, but was done so for the sake of 
expediently correcting a significant systematic error. Instead, in future, 
it would be better for all 
studies that fit models to number counts of bulge RCG stars to first correct 
for incompleteness using artificial star tests. Without doing so it is 
possible that models fit to the magnitude distribution around the red clump 
might systematically underestimate the number of stars, as well as the 
location of the clump if the incompleteness varies as a function of magnitude (which 
it almost certainly does). Failure to do so could impact studies of galactic 
structure~\citep[e.g.][]{Rattenbury2007,Cao2013,Wegg2013}, 
interstellar extinction~\citep{sumEX04,Nataf2013, Gonzalez2011} and as we have shown here, 
microlensing event rates and optical depths.

\acknowledgments
TS acknowledges the financial support from the JSPS, JSPS23340044, JSPS24253004. Work by MTP was performed under contract with the California Institute of Technology (Caltech)/Jet Propulsion Laboratory (JPL) funded by NASA through the Sagan Fellowship Program executed by the NASA Exoplanet Science Institute.


\clearpage



\begin{deluxetable}{lrrrrrrrrrc}
\tabletypesize{\scriptsize}
\tablecaption{MOA-II Galactic bulge fields with Galactic coordinates of the mean field center ($<l>$, $<b>$), 
the number of subfields used ($N_{\rm sub}$), the number of frames ($N_{\rm f}$), 
the number of source stars ($N_{\rm s}$ in thousands),
the number of microlensing events ($N_{\rm ev}$), 
the microlensing event rate per star per year ($\Gamma$),
the microlensing event rate per square degree per year ($\Gamma_{\rm deg^2}$), 
the optical depth ($\tau_{200}$), 
and the mean detection efficiency weighted $t_{\rm E}$. \label{tbl:fld}}
\tablewidth{0pt}
\tablehead{
    \colhead{Field} & \colhead{$<l>$} & \colhead{$<b>$} & \colhead{$N_{\rm sub}$}  
  & \colhead{$N_{\rm f}$}  & \colhead{$N_{\rm s}$} & \colhead{$N_{\rm ev}$} 
  & \colhead{$\tau_{200}$} & \colhead{$\Gamma (10^{-6})$} &  \colhead{$\Gamma_{\rm deg^2}$} & \colhead{$<t_{\rm E}>$} \\
    \colhead{} & \colhead{($^\circ$)} & \colhead{($^\circ$)} & \colhead{}  
  & \colhead{} & \colhead{$(10^3)$}  & \colhead{}
  & \colhead{(10$^{-6}$)} & \colhead{(star$^{-1}$yr$^{-1}$)} &  \colhead{(deg.$^{-2}$yr$^{-1}$)}  & \colhead{(day)}
}
\startdata
 gb1 & -4.3306 & -3.1119 &   79 & 2253 &    5356 &  22 &   2.03$_{ -0.69}^{+  1.53}$ & 15.0$_{-3.3}^{+4.2}$ & 37.0$_{-8.1}^{+10.4}$ & 30.6 \\
 gb2 & -3.8624 & -4.3936 &   79 & 2386 &    5527 &  17 &   0.80$_{ -0.19}^{+  0.25}$ & 9.7$_{-2.1}^{+2.7}$ & 24.7$_{-5.4}^{+6.9}$ & 18.8 \\
 gb3 & -2.3463 & -3.5133 &   79 & 2067 &    5635 &  19 &   1.07$_{ -0.26}^{+  0.34}$ & 9.1$_{-1.9}^{+2.4}$ & 23.5$_{-4.8}^{+6.2}$ & 26.7 \\
 gb4 & -0.8210 & -2.6317 &   77 & 2985 &    5506 &  41 &   2.31$_{ -0.46}^{+  0.58}$ & 27.0$_{-4.6}^{+5.6}$ & 70.2$_{-11.9}^{+14.7}$ & 19.4 \\
 gb5 &  0.6544 & -1.8595 &   65 & 8229 &    6106 &  67 &   2.81$_{ -0.39}^{+  0.45}$ & 36.6$_{-4.4}^{+4.9}$ & 124.9$_{-15.1}^{+16.7}$ & 17.4 \\
 gb6 &  1.8405 & -1.4890 &   11 & 1779 &     446 &   4 &   1.38$_{ -0.57}^{+  0.90}$ & 18.5$_{-7.4}^{+12.3}$ & 27.3$_{-10.9}^{+18.2}$ & 16.9 \\
 gb7 & -1.7147 & -4.5992 &   78 & 1970 &    5082 &  15 &   0.70$_{ -0.18}^{+  0.25}$ & 8.4$_{-2.0}^{+2.6}$ & 20.0$_{-4.6}^{+6.1}$ & 18.8 \\
 gb8 & -0.1937 & -3.7495 &   78 & 2139 &    6366 &  16 &   0.71$_{ -0.17}^{+  0.22}$ & 7.4$_{-1.7}^{+2.1}$ & 22.1$_{-4.9}^{+6.3}$ & 21.7 \\
 gb9 &  1.3329 & -2.8786 &   79 & 8301 &    9881 &  74 &   2.59$_{ -0.39}^{+  0.47}$ & 26.5$_{-3.1}^{+3.5}$ & 120.4$_{-14.0}^{+15.9}$ & 22.2 \\
gb10 &  2.8448 & -2.0903 &   70 & 1992 &    4978 &  36 &   3.60$_{ -0.92}^{+  1.23}$ & 28.9$_{-5.1}^{+6.4}$ & 74.6$_{-13.3}^{+16.6}$ & 28.3 \\
gb11 & -1.1093 & -5.7257 &   76 & 2004 &    4023 &   8 &   0.44$_{ -0.14}^{+  0.21}$ & 5.9$_{-2.0}^{+3.0}$ & 11.3$_{-3.8}^{+5.7}$ & 16.9 \\
gb12 &  0.4391 & -4.8658 &   79 & 1790 &    5510 &  12 &   0.94$_{ -0.30}^{+  0.43}$ & 6.4$_{-1.6}^{+2.1}$ & 16.2$_{-4.1}^{+5.4}$ & 33.2 \\
gb13 &  1.9751 & -4.0190 &   79 & 1811 &    8133 &  27 &   1.76$_{ -0.48}^{+  0.64}$ & 13.5$_{-2.5}^{+3.1}$ & 50.7$_{-9.3}^{+11.5}$ & 29.6 \\
gb14 &  3.5083 & -3.1698 &   79 & 1770 &    7934 &  29 &   1.58$_{ -0.42}^{+  0.57}$ & 14.0$_{-2.4}^{+2.9}$ & 51.2$_{-8.8}^{+10.8}$ & 25.5 \\
gb15 &  4.9940 & -2.4496 &   62 & 1952 &    2448 &  14 &   1.77$_{ -0.47}^{+  0.64}$ & 17.1$_{-4.1}^{+5.3}$ & 24.6$_{-5.8}^{+7.6}$ & 23.5 \\
gb16 &  2.6048 & -5.1681 &   79 & 1756 &    5627 &  17 &   1.40$_{ -0.38}^{+  0.50}$ & 9.0$_{-2.0}^{+2.5}$ & 23.3$_{-5.1}^{+6.5}$ & 35.3 \\
gb17 &  4.1498 & -4.3365 &   79 & 1792 &    6448 &  16 &   1.14$_{ -0.30}^{+  0.41}$ & 8.2$_{-1.9}^{+2.4}$ & 24.3$_{-5.5}^{+7.1}$ & 31.7 \\
gb18 &  5.6867 & -3.5055 &   78 & 1799 &    4722 &  13 &   0.82$_{ -0.24}^{+  0.33}$ & 7.8$_{-2.0}^{+2.6}$ & 17.1$_{-4.3}^{+5.6}$ & 23.9 \\
gb19 &  6.5534 & -4.5749 &   78 & 1704 &    4424 &  12 &   0.94$_{ -0.26}^{+  0.36}$ & 7.1$_{-1.8}^{+2.4}$ & 14.6$_{-3.7}^{+4.9}$ & 30.0 \\
gb20 &  8.1025 & -3.7531 &   79 & 1679 &    3673 &  12 &   0.97$_{ -0.27}^{+  0.37}$ & 8.9$_{-2.2}^{+3.0}$ & 15.1$_{-3.8}^{+5.0}$ & 24.6 \\
gb21 &  9.6172 & -2.9318 &   73 & 1659 &    2419 &   3 &   0.26$_{ -0.11}^{+  0.23}$ & 3.5$_{-1.7}^{+2.6}$ & 4.2$_{-2.0}^{+3.1}$ & 17.0 \\
all &  1.8530 & -3.6890 & 1536 & --- &  110253 & 474 &   1.53$_{ -0.11}^{+  0.12}$ & 14.5$_{-0.7}^{+0.7}$ & 37.8$_{-1.9}^{+1.9}$ & 24.0 \\
all$_{\rm RC}^*$  &  1.8530 & -3.6890 & 1536 & --- &    7997 &  83 &   1.28$_{ -0.19}^{+  0.22}$ & 15.1$_{-1.6}^{+1.8}$ & 2.9$_{-0.3}^{+0.3}$ & 19.2 \\
\enddata
\tablecomments{
The values are for the all-source sample except for all$_{\rm RCG}$ which is for the RCG source sample.}
\end{deluxetable}

\clearpage

\begin{deluxetable}{rrrrrrr}
\tabletypesize{\footnotesize}
\tablecaption{Microlensing optical depth and event rates binned in $b$ for the all-source
sample with $|l|<5^\circ$.
 \label{tbl:opt_binb_all_lth5}}
\tablewidth{0pt}
\tablehead{
\colhead{$<b>^*$} &
\colhead{$N_{\rm sub}$} &
\colhead{$N_{\rm s}$} &
\colhead{$N_{\rm ev}$} &
\colhead{$\tau (10^{-6})$} &
\colhead{$\Gamma$\,$(10^{-6})$} &
\colhead{$\Gamma_{\rm deg^2}$} \\
\colhead{$(^\circ)$} &
\colhead{$$} &
\colhead{} &
\colhead{$$} &
\colhead{$$} &
\colhead{(star$^{-1}$ yr$^{-1}$)} &
\colhead{(deg.$^{-2}$yr$^{-1}$)} 
}
\startdata
  -1.4012 &  20  &   687319 &  12 & 3.14$_{ -0.85}^{+1.19}$ &   50.0$_{-13.1}^{+ 17.7}$ &   62.4$_{-16.3}^{+ 22.1}$\\
  -1.7690 &  70  &  5032788 &  52 & 3.61$_{ -0.66}^{+0.81}$ &   34.8$_{ -4.8}^{+  5.5}$ &   90.9$_{-12.6}^{+ 14.5}$\\
  -2.2645 & 114  &  9056629 &  70 & 2.61$_{ -0.49}^{+0.61}$ &   30.7$_{ -4.1}^{+  4.8}$ &   88.6$_{-11.7}^{+ 13.7}$\\
  -2.7576 & 146  & 13187560 &  75 & 2.58$_{ -0.53}^{+0.68}$ &   20.9$_{ -2.4}^{+  2.7}$ &   68.8$_{ -7.9}^{+  9.0}$\\
  -3.2486 & 168  & 15542979 &  67 & 1.50$_{ -0.28}^{+0.34}$ &   15.0$_{ -1.9}^{+  2.1}$ &   50.6$_{ -6.3}^{+  7.0}$\\
  -3.7490 & 172  & 14776708 &  58 & 1.26$_{ -0.19}^{+0.22}$ &   12.9$_{ -1.7}^{+  1.9}$ &   40.3$_{ -5.2}^{+  5.8}$\\
  -4.2512 & 172  & 13727488 &  43 & 1.25$_{ -0.22}^{+0.27}$ &    9.8$_{ -1.5}^{+  1.7}$ &   28.6$_{ -4.2}^{+  4.9}$\\
  -4.7410 & 154  & 10977355 &  22 & 0.67$_{ -0.15}^{+0.20}$ &    5.8$_{ -1.1}^{+  1.4}$ &   15.0$_{ -2.9}^{+  3.6}$\\
  -5.2270 & 101  &  6558015 &  16 & 0.85$_{ -0.25}^{+0.35}$ &    6.9$_{ -1.5}^{+  2.0}$ &   16.2$_{ -3.6}^{+  4.6}$\\
  -5.7197 &  56  &  3099616 &   8 & 1.25$_{ -0.48}^{+0.78}$ &    6.6$_{ -2.0}^{+  2.9}$ &   13.4$_{ -4.0}^{+  5.8}$\\
  -6.2282 &  21  &  1030160 &   4 & 0.81$_{ -0.35}^{+0.58}$ &   13.3$_{ -5.8}^{+ 10.9}$ &   23.7$_{-10.4}^{+ 19.5}$\\
\enddata
\tablecomments{
$*$Average galactic latitude of fields in each bin.
$N_{\rm sub}$, $N_{\rm s}$ and $N_{\rm ev}$ indicate the number of subfields, source stars and microlensing events
in each bin.
}
\end{deluxetable}

\begin{deluxetable}{rrrrrrrrr}
\tabletypesize{\footnotesize}
\tablecaption{Microlensing optical depth and event rates binned in $b$ for the RCG
sample with $|l|<5^\circ$.
 \label{tbl:opt_binb_RCG_lth5}}
\tablewidth{0pt}
\tablehead{
\colhead{$<b>^*$} &
\colhead{$N_{\rm sub}$} &
\colhead{$N_{\rm s}$} &
\colhead{$N_{\rm ev}$} &
\colhead{$\tau (10^{-6})$} &
\colhead{$\Gamma$\,$(10^{-6})$} &
\colhead{$\Gamma_{\rm deg^2}$} \\
\colhead{$(^\circ)$} &
\colhead{$$} &
\colhead{} &
\colhead{$$} &
\colhead{$$} &
\colhead{(star$^{-1}$ yr$^{-1}$)} &
\colhead{(deg.$^{-2}$yr$^{-1}$)} 
}
\startdata
  -1.6872 &  90  &   715368 &  16 & 2.06$_{ -0.54}^{+0.74}$ &   33.9$_{ -7.6}^{+  9.7}$ &    9.8$_{ -2.2}^{+  2.8}$\\
  -2.2645 & 114  &   807674 &  16 & 2.57$_{ -0.77}^{+1.08}$ &   28.9$_{ -6.5}^{+  8.3}$ &    7.5$_{ -1.7}^{+  2.2}$\\
  -2.7576 & 146  &   976651 &  11 & 1.08$_{ -0.35}^{+0.52}$ &   16.2$_{ -4.3}^{+  5.7}$ &    3.9$_{ -1.0}^{+  1.4}$\\
  -3.2486 & 168  &  1051602 &  14 & 1.54$_{ -0.44}^{+0.60}$ &   18.8$_{ -4.5}^{+  5.9}$ &    4.3$_{ -1.0}^{+  1.3}$\\
  -3.7490 & 172  &   952935 &  11 & 1.61$_{ -0.53}^{+0.81}$ &   17.5$_{ -4.7}^{+  6.4}$ &    3.5$_{ -0.9}^{+  1.3}$\\
  -4.2512 & 172  &   863385 &   4 & 0.48$_{ -0.21}^{+0.48}$ &    6.2$_{ -2.4}^{+  4.3}$ &    1.1$_{ -0.4}^{+  0.8}$\\
  -5.1480 & 332  &  1368724 &   3 & 0.83$_{ -0.37}^{+0.88}$ &    4.2$_{ -2.1}^{+  3.5}$ &    0.6$_{ -0.3}^{+  0.5}$\\
\enddata
\tablecomments{
$*$Average galactic latitude of fields in each bin. The notation is the same as in Table \ref{tbl:opt_binb_all_lth5}.
}
\end{deluxetable}

\begin{deluxetable}{lccrrrrrr}
\tabletypesize{\footnotesize}
\tablecaption{Average microlensing optical depth and event rates at the position of each
subfield for the all-source sample.
 \label{tbl:opt_2D}}
\tablewidth{0pt}
\tablehead{
\colhead{subfield} &
\colhead{$l$} &
\colhead{$b$} &
\colhead{$N_{\rm sub}$} &
\colhead{$N_{\rm s}$} &
\colhead{$N_{\rm ev}$} &
\colhead{$\tau (10^{-6})$} &
\colhead{$\Gamma$ ($10^{-6}$)} &
\colhead{$\Gamma_{\rm deg^2}$}  \\
\colhead{} &
\colhead{$(^\circ)$} &
\colhead{$(^\circ)$} &
\colhead{} &
\colhead{} &
\colhead{} &
\colhead{} &
\colhead{(star$^{-1}$yr$^{-1}$)} &
\colhead{(deg.$^{-2}$yr$^{-1})$} 
}
\startdata
   gb5-1-3 &  1.1704 & -1.3459 &  58 & 4796785 & 55 & $ 2.9_{ -0.5}^{+ 0.7}$ & $ 42.4_{-10.2}^{+12.9}$ & $109.0_{-26.3}^{+ 33.1}$\\
   gb5-1-7 &  1.3125 & -1.2630 &  51 & 3917583 & 46 & $ 2.9_{ -0.6}^{+ 0.7}$ & $ 42.4_{-10.9}^{+15.3}$ & $ 96.6_{-24.8}^{+ 34.9}$\\
   gb5-2-2 &  0.7835 & -1.3776 &  59 & 5177353 & 59 & $ 2.7_{ -0.5}^{+ 0.5}$ & $ 40.2_{ -9.1}^{+11.7}$ & $118.0_{-26.7}^{+ 34.5}$\\
   gb5-2-3 &  0.8685 & -1.5224 &  69 & 6310401 & 67 & $ 2.7_{ -0.4}^{+ 0.5}$ & $ 40.2_{ -7.9}^{+ 9.9}$ & $126.1_{-24.8}^{+ 31.1}$\\
   gb5-2-6 &  0.9280 & -1.2935 &  54 & 4439824 & 49 & $ 2.8_{ -0.5}^{+ 0.6}$ & $ 41.7_{-10.1}^{+13.4}$ & $112.9_{-27.4}^{+ 36.1}$\\
   gb5-2-7 &  1.0130 & -1.4379 &  63 & 5557000 & 63 & $ 2.8_{ -0.5}^{+ 0.6}$ & $ 41.7_{ -8.7}^{+11.2}$ & $120.2_{-25.2}^{+ 32.1}$\\
   gb5-3-1 &  0.3942 & -1.4104 &  53 & 5008885 & 52 & $ 2.7_{ -0.5}^{+ 0.6}$ & $ 34.2_{ -8.7}^{+11.4}$ & $103.0_{-26.1}^{+ 34.2}$\\
   gb5-3-2 &  0.4788 & -1.5549 &  67 & 6338841 & 63 & $ 2.7_{ -0.5}^{+ 0.6}$ & $ 35.1_{ -7.3}^{+ 9.3}$ & $114.3_{-23.7}^{+ 30.2}$\\
   gb5-3-3 &  0.5639 & -1.6998 &  78 & 7640959 & 74 & $ 2.7_{ -0.4}^{+ 0.5}$ & $ 35.9_{ -6.9}^{+ 8.3}$ & $126.2_{-24.1}^{+ 29.1}$\\
   gb5-3-6 &  0.6239 & -1.4697 &  65 & 5938827 & 65 & $ 2.7_{ -0.4}^{+ 0.5}$ & $ 37.8_{ -8.1}^{+10.5}$ & $117.9_{-25.4}^{+ 32.8}$\\
   gb5-3-7 &  0.7091 & -1.6146 &  76 & 7142733 & 75 & $ 2.7_{ -0.4}^{+ 0.5}$ & $ 38.4_{ -7.2}^{+ 9.0}$ & $128.6_{-24.0}^{+ 30.2}$\\
   gb5-4-0 &  0.0089 & -1.4439 &  48 & 4203315 & 37 & $ 2.5_{ -0.6}^{+ 0.7}$ & $ 28.8_{ -8.5}^{+11.3}$ & $ 82.1_{-24.2}^{+ 32.2}$\\
   gb5-4-1 &  0.0918 & -1.5877 &  59 & 5398262 & 52 & $ 2.7_{ -0.5}^{+ 0.7}$ & $ 29.6_{ -7.0}^{+ 9.4}$ & $ 92.6_{-21.9}^{+ 29.2}$\\
   gb5-4-2 &  0.1755 & -1.7322 &  73 & 6845398 & 65 & $ 2.8_{ -0.5}^{+ 0.6}$ & $ 30.5_{ -6.5}^{+ 7.9}$ & $104.0_{-22.1}^{+ 27.0}$\\
   gb5-4-3 &  0.2599 & -1.8771 &  81 & 7762864 & 70 & $ 2.8_{ -0.5}^{+ 0.6}$ & $ 31.3_{ -5.9}^{+ 7.2}$ & $114.7_{-21.5}^{+ 26.5}$\\
   gb5-4-5 &  0.2356 & -1.5028 &  56 & 5313798 & 53 & $ 2.7_{ -0.5}^{+ 0.7}$ & $ 31.6_{ -7.8}^{+10.1}$ & $ 97.5_{-24.2}^{+ 31.1}$\\
   gb5-4-6 &  0.3197 & -1.6474 &  70 & 6687816 & 67 & $ 2.7_{ -0.5}^{+ 0.6}$ & $ 32.7_{ -6.8}^{+ 8.6}$ & $110.1_{-22.8}^{+ 28.9}$\\
   gb5-4-7 &  0.4044 & -1.7925 &  81 & 7829673 & 76 & $ 2.8_{ -0.5}^{+ 0.5}$ & $ 33.7_{ -6.6}^{+ 7.9}$ & $121.9_{-23.9}^{+ 28.4}$\\
   gb5-5-0 & -0.2872 & -1.6227 &  53 & 4406168 & 43 & $ 2.6_{ -0.5}^{+ 0.7}$ & $ 29.2_{ -8.9}^{+12.7}$ & $ 81.0_{-24.8}^{+ 35.3}$\\
   gb5-5-1 & -0.2055 & -1.7661 &  63 & 5424411 & 50 & $ 2.8_{ -0.6}^{+ 0.7}$ & $ 28.7_{ -7.6}^{+10.1}$ & $ 86.8_{-22.9}^{+ 30.6}$\\
   gb5-5-2 & -0.1227 & -1.9100 &  76 & 6720615 & 57 & $ 2.9_{ -0.6}^{+ 0.7}$ & $ 28.5_{ -6.8}^{+ 9.0}$ & $ 93.1_{-22.4}^{+ 29.5}$\\
\enddata
\tablecomments{
The averages include all the subfields within $1^\circ$ of the center of each subfield with
a Gaussian weighting function with $\sigma = 0^\circ .4$.
$N_{\rm sub}$, $N_{\rm s}$ and $N_{\rm ev}$ are numbers of subfields, source stars and microlensing events 
in this $1^\circ$ circle, respectively. 
A complete electronic version of this table is available at http://iral2.ess.sci.osaka-u.ac.jp/\~{}sumi/OPTMOAII\_Nataf/Table4.dat
}
\end{deluxetable}
\begin{deluxetable}{lr}
\tablecaption{The best 2D model parameters for  $\Gamma$. 
 \label{tbl:param_2D}}
\tablewidth{0pt}
\tablehead{
\colhead{param} &
\colhead{value} \\
}
\startdata
$ a_{ 0} $ &    76.558396  \\
$ a_{ 1} $ &     0.758556  \\
$ a_{ 2} $ &    32.598859 \\
$ a_{ 3} $ &    -0.274198 \\
$ a_{ 4} $ &     0.178113  \\
$ a_{ 5} $ &     4.408679  \\
$ a_{ 6} $ &    -0.017363 \\
$ a_{ 7} $ &    -0.104587 \\
$ a_{ 8} $ &    -0.006764  \\
$ a_{ 9} $ &     0.157305  \\
$ a_{10} $ &     0.651233 \\
$ a_{11} $ &    -0.717574 \\
$ a_{12} $ &     0.163776  \\
$ a_{13} $ &     0.324459  \\
$ a_{14} $ &     0.005950  \\
$ a_{15} $ &     0.032564  \\
\enddata
\tablecomments{  The model parameters are defined in Equation (\ref{eq:2Dformula}).
}
\end{deluxetable}


\begin{thebibliography}{}

\bibitem[Afonso et al.(2003)]{afo03}Afonso, C. et al. 2003, A\&A, 404, 145
\bibitem[Alard(2000)]{ala00}Alard C., 2000, A\&AS, 144, 363
\bibitem[Alard \& Lupton(1998)]{ala98}Alard C., Lupton R. H., 1998, ApJ, 503, 325
\bibitem[Alcock et al.(1997)]{alc97}Alcock, C. et al. 1997, ApJ, 486, 697
\bibitem[Alcock et al.(2000)]{alc00b}Alcock C. et al., 2000b, ApJ, 541, 734
\bibitem[Awiphan, Kerins \& Robin(2016)]{Awiphan2016} Awiphan, S., Kerins, E. \& Robin, A. C., 2016, MNRAS, 456, 1666
\bibitem[Bennett \& Rhie(2002)]{ben02}Bennett, D.P. \& Rhie, S.H.\ 2002, \apj, 574, 985
\bibitem[Bensby et al.(2013)]{Bensby2013}Bensby, T., Yee, J. C., Feltzing, S., et al. 2013, A\&A, 549, A147
\bibitem[Bissantz \& Gerhard(2002)]{bis02}Bissantz, N. \& Gerhard, O. 2002, MNRAS, 330, 591
\bibitem[Bond et al.(2001)]{bon01}Bond I. A. et al., 2001, MNRAS, 327, 868 
\bibitem[Cao et al.(2013)]{Cao2013}Cao, L., Mao, S., Nataf, D., Rattenbury, N. J., \& Gould, A., 2013, MNRAS, 434, 595 
\bibitem[Evans \& Belokurov(2002)]{eva02} Evans N.W., \& Belokurov, 2002, ApJ, 567, 119 
\bibitem[Gonzalez et al.(2011)]{Gonzalez2011}Gonzalez, O. A., Rejkuba, M., Zoccali, M., Valenti, E., \& Minniti, D. 2011, \aap, 534, A3
\bibitem[Gould \& An(2002)]{gou02}Gould, A. \& An, J. H. 2002, ApJ, 565, 1381
\bibitem[Green et al.(2012)]{green2012}Green, J. et al., 2012, preprint, astro-ph/1208.4012
\bibitem[Griest et al.(1991)]{gri91}Griest, K., et al. 1991, ApJ, 372, L79
\bibitem[Gyuk(1999)]{gyu99}Gyuk, G. 1999, ApJ, 510, 205 
\bibitem[Hamadache et al.(2006)]{ham06} Hamadache, C., Le Guillou, L., Tisserand, P., et al.\ 2006, \aap, 454, 185
\bibitem[Han \& Gould(1995)]{han95}Han, C. \& Gould, A. 1995, ApJ, 449, 521
\bibitem[Han \& Gould(2003)]{han03}Han, C. \& Gould, A. 2002, ApJ, 592, 172
\bibitem[Han(1999)]{han99}Han, C. 1999, MNRAS, 309, 373
\bibitem[Holtzman et al.(1998)]{hol98}Holtzman, J.~A., Watson, A.~M., Baum, W.~A., et al.\ 1998, \aj, 115, 1946
\bibitem[Kerins, Robin \& Marshal(2009)]{kerins2009} Kerins, E., Robin, A. C., \& Marshal, D. J. 2009, MNRAS, 396, 1202
\bibitem[Kim et al.(2010)]{kmtnet} Kim, S.-L., Park, B.-G., Lee, C.-U., et al.\ 2010, \procspie, 7733, 77333F
\bibitem[Kiraga \& Paczy\'{n}ski(1994)]{kir94}Kiraga, M., \& Paczy\'{n}ski, B. 1994, ApJ, 430, L101
\bibitem[Kiraga, Paczy\'{n}ski \& Stanek(1997)]{kir97}Kiraga, M., Paczy\'{n}ski, B.  \& Stanek, K. Z.,  1997, ApJ, 485, 611
\bibitem[Nataf et al.(2013)]{Nataf2013}Nataf, D. M. et al. 2013,  \apj, 769, 88
\bibitem[Novati et al.(2008)]{novati2008}Novati S.C., Luca, F. De., Jetzer, Ph.,  Mancini, L., \& Scarpetta, G.  2008, A\&A, 480, 723
\bibitem[Paczy\'{n}ski(1986)]{pac86}Paczy\'{n}ski, B.  1986, ApJ, 304, 1 
\bibitem[Paczy\'{n}ski(1991)]{pac91}Paczy\'{n}ski, B.  1991, ApJ, 371, L63
\bibitem[Paczy\'{n}ski et al.(1994)]{pac94}Paczy\'{n}ski, B. et al. 1994, ApJ., 435, L113
\bibitem[Paczy\'{n}ski(1996)]{pac96}Paczy\'{n}ski, B. 1996, ARA\&A, 34, 419
\bibitem[Peale(1998)]{pea98}Peale, S. J. 1998, ApJ, 509, 177
\bibitem[Penny et al.(2013)]{penny13} Penny, M.~T., Kerins, E., Rattenbury, N. J., et al.\ 2013,  \mnras, 434, 2
\bibitem[Popowski et al.(2001)]{pop01}Popowski, P. et al. 2001, in ASP Conference Series: Microlensing 2000: A New Era of Microlensing Astrophysics, eds. J.W. Menzies \& P.D. Sackett (San Francisco: Astronomical Society of the Pacific), Vol. 239, p. 244, (astro-ph/0005466)
\bibitem[Popowski et al.(2005)]{pop05}Popowski, P. et al. 2005, \apj, 631, 879
\bibitem[Rattenbury et al.(2007)]{Rattenbury2007} Rattenbury, N.J., Mao, S., Sumi, T., \& Smith, M. C.\ 2007,  \mnras, 378, 1064
\bibitem[Sako et al.(2008)]{sako2008}Sako, T., et al. 2008, Experimental Astronomy, 22, 51
\bibitem[Schechter, Mateo \& Saha(1993)]{sch93}Schechter, L., Mateo, M., \& Saha, A., 1993, PASP, 105, 1342S
\bibitem[Shvartzvald \& Maoz(2012)]{wise_survey} Shvartzvald, Y., \& Maoz, D.\ 2012, \mnras, 419, 3631 
\bibitem[Smith, Wo\'{z}niak, Mao \& Sumi(2007)]{smith2007}Smith, M. C., Wo\'{z}niak, P. R., Mao, S. \& Sumi, T., 2007, MNRAS, 380, 805
\bibitem[Spergel et al.(2015)]{Spergel2015}Spergel, D. et al., 2015, preprint, astro-ph/1503.03757
\bibitem[Stanek et al.(2000)]{sta00}Stanek, K. Z. et al. 2000, Acta Astronomica, 50, 191
\bibitem[Sumi(2004)]{sumEX04}Sumi, T., 2004, MNRAS, 349, 193
\bibitem[Sumi et al.(2003)]{sumi03}Sumi, T. et al., 2003, ApJ, 591, 204  
\bibitem[Sumi et al.(2006)]{sumi2006}Sumi, T. et al., 2006, ApJ, 636, 240
\bibitem[Sumi et al.(2011)]{sumi2011}Sumi, T. et al., 2011, Nature, 473, 349
\bibitem[Sumi et al.(2013)]{sumi2013}Sumi, T. et al., 2013, ApJ, 778, 150
\bibitem[Szyma\'{n}ski et al.(2011)]{Szymanski2011}Szyma\'{n}ski, M. K., Udalski, A., Soszy\'{n}ski, I., et al. 2011, AcA, 61, 83
\bibitem[Tomany \& Crotts(1996)]{tom96}Tomany, A. B. \& Crotts, A. P., 1996, AJ, 112, 2872
\bibitem[Udalski et al.(1994)]{uda94}Udalski, A. et al. 1994, Acta Astronomica, 44, 165
\bibitem[Udalski et al.(2002)]{uda02}Udalski A. et al. 2002, Acta Astronomica, 52, 217
\bibitem[Udalski(2003)]{uda03}Udalski, A. 2003, Acta Astronomica, 53, 291
\bibitem[Wegg  \& Gerhard(2013)]{Wegg2013} Wegg, C. \& Gerhard, O. 2013,  \mnras, 435, 1874 
\bibitem[Wood \& Mao(2005)]{wood05} Wood, A., \& Mao, S. 2005,  \mnras, 362, 945 
\bibitem[Wo\'{z}niak \& Paczy\'{n}ski(1997)]{woz97}Wo\'{z}niak P. R., \& Paczy\'{n}ski, B. 1997, ApJ, 487, 55
\bibitem[Wo\'{z}niak et al.(2001)]{woz01}Wo\'{z}niak, P. R., et al. 2001, Acta Astronomica, 51, 175
\bibitem[Zhao \& Mao(1996)]{zha96}Zhao, H. \& Mao, S. 1996, MNRAS, 283, 1197
\bibitem[Zhao, Spergel \& Rich(1995)]{zha95}Zhao, H., Spergel, D. N. \& Rich, R. 1995, ApJ, 440, L13
\end{thebibliography}
\end{document}